\documentstyle[epsfig, subfigure, lscape]{mn}

\def\ltapprox{\raise 2pt \hbox {$<$} \kern-1.1em \lower 5pt \hbox {$\approx$}}
\def\ltsim{\raise 2pt \hbox {$<$} \kern-1.1em \lower 4pt \hbox {$\sim$}}
\def\gtsim{\raise 2pt \hbox {$>$} \kern-1.1em \lower 4pt \hbox {$\sim$}}

\title{Cluster mergers and non-thermal phenomena : a
statistical magneto--turbulent model}
\author[R. Cassano \& G. Brunetti]
       {R. Cassano,$^{1,2}$ 
        G. Brunetti$^{2}$\\
       $^1$ Dipartimento di Astronomia,Universita' di 
       Bologna, via Ranzani 1, I-40127 Bologna, Italy\\
       $^2$ Istituto di Radioastronomia del CNR, via Gobetti 101,
       I--40129 Bologna, Italy\\
}

\begin{document}
\maketitle

\begin{abstract}

There is now firm evidence that the ICM 
consists of a mixture of hot plasma, magnetic 
fields and relativistic particles. 
The most important evidences for non-thermal phenomena
in galaxy clusters comes from the spectacular
synchrotron radio emission diffused over Mpc scale 
observed in a growing number of massive 
clusters and, more recently, in the hard X--ray tails detected
in a few cases in excess of the thermal bremmstrahlung
spectrum.
A promising possibility to explain giant radio halos
is given by the presence of relativistic electrons
reaccelerated by some kind of turbulence generated
in the cluster volume during merger events.
With the aim to investigate the connection between
thermal and non--thermal properties of the ICM, 
in this paper we develope a statistical magneto-turbulent 
model which describes in a self-consistent way the evolution 
of the thermal ICM and that of the non-thermal emission from clusters.
Making use of the extended Press \& Schechter formalism, 
we follow cluster mergers and estimate the injection rate of
the fluid turbulence generated during these energetic events.
We then calculate the evolution of the spectrum of
the relativistic electrons in the ICM 
during the cluster life by taking into account both
the electron--acceleration due to the merger--driven turbulence and
the relevant energy losses of the electrons.
We end up with a synthetic population of 
galaxy clusters for which 
the evolution of the ICM and of the non--thermal spectrum
emitted by the accelerated electrons is calculated.
The generation of detectable non--thermal radio and hard X--ray
emission in the simulated clusters is found to be possible during
major merger events for reliable values of the model parameters.
In addition the occurrence of radio halos as a function
of the mass of the parent clusters is calculated and compared with
observations. In this case it is found that the model expectations
are in good agreement with observations: radio halos are 
found in about 30\% of the more massive clusters in our synthetic 
population ($M$ $\gtsim$ $1.8 \times 10^{15}$ M$_{\odot}$) and in about 
4\% of the intermediate massive clusters 
($9 \times 10^{14}< M < 1.8\times 10^{15}M_{\odot}$), while the radio
halo phenomenon is found to be extremely rare in the case of
the smaller clusters.
\end{abstract}

\begin{keywords}
acceleration of particles - turbulence - radiation mechanisms: non--thermal -
galaxies: clusters: general - 
radio continuum: general - X--rays: general
\end{keywords}

\maketitle

\section{Introduction}

\subsection{Evidence for non-thermal phenomena}

There is now firm evidence that the ICM is a mixture of hot gas, magnetic 
fields and relativistic particles. 

The most important evidence for relativistic electrons in clusters
of galaxies comes from the diffuse synchrotron radio emission observed in
a growing number of massive clusters.
The diffuse emissions are referred to as radio halos and/or radio
mini--halos when they appear confined to the center of the
cluster, while they are called relics when they are found in the
cluster periphery (e.g., Feretti, 2003).

Diffuse radio emission is not the only evidence of non-thermal 
activity in the ICM. 
Additional evidence, comes from the detection of 
hard X-ray (HXR) excess emission in the case of the Coma cluster, A2256 
and possibly A754 (Fusco--Femiano et 
al., 1999, 2000,03,04; 
Rephaeli et al., 1999; Rephaeli \& Gruber, 2002,03). 
If these excesses are indeed of non-thermal origin,
they may be explained in terms of IC scattering of relativistic
electrons off the photons of the cosmic microwave background (CMB)
(Fusco--Femiano et al., 1999,2000; Rephaeli et al., 1999; V\"olk \& 
Atoyan 1999; Brunetti et al.2001a; Petrosian 2001; Fujita \& Sarazin 2001;
Blasi 2001).
Unfortunately the poor sensitivity of the present
and past HXR-facilities does not allow to obtain a 
iron-clad detection of these excesses
and thus future observatories (e.g. ASTRO-E2, NEXT) are necessary 
to definitely confirm them (see Rossetti \& Molendi 2004;
Fusco-Femiano et al.2004), to constrain their spectral shape, and
hopefully to increase the statistics of HXR excesses.

The presence of high
energy hadrons is at present not yet proven, but in principle, due to 
confinement of cosmic rays over cosmological time scales 
(e.g., V\"{o}lk et al. 1996; Berezinsky, Blasi \& Ptuskin 1997), the
hadron content of the intracluster medium might be appreciable
and may be constrained by future gamma--ray observations
(e.g., Blasi 2003; Miniati 2003).

\subsection{Thermal -- non-thermal connection}

Giovannini, Tordi and Feretti (1999) found that $\sim$5\% of clusters 
in a complete X-ray flux limited sample 
(from Ebeling et al.1996) have a radio halo source. 
The detection rate of radio halos shows an abrupt increase
with increasing the X-ray luminosity of the host clusters.
Indeed it has been found
that about 30-35\% of the galaxy clusters with 
X-ray luminosity larger than 10$^{45}$ erg s$^{-1}$
show diffuse non-thermal radio emission
(Giovannini \& Feretti 2002).
The high luminosity of clusters hosting radio halos implies 
that these clusters also have a high temperature 
($kT$ \gtsim $7$ keV) and
a large mass (\gtsim $10^{15}$ M$_{\odot}$).
Although still based on relatively poor statistics, 
these observations suggest a leading role of
the cluster mass and/or temperature
in the formation of radio halos.

In addition correlation between the radio 
power of radio halos at 1.4 GHz and 
the bolometric X--ray luminosity, temperature
and mass of the parent clusters have been found (Colafrancesco 1999;
Liang et al. 2000; Giovannini \& Feretti 2002; Govoni et al.2001).

\subsection{Models}

The difficulty in explaining the extended radio halos arises from the 
combination of their $\sim$Mpc size, and the relatively short radiative 
lifetime of the radio emitting electrons. Indeed, the diffusion time 
necessary for the radio electrons to cover such distances is orders of 
magnitude larger than their radiative lifetime.
As proposed first by Jaffe (1977), a solution to this puzzle would be
provided by continuous {\it in situ} reacceleration of the relativistic 
electrons. 

An alternative to the reacceleration scenario is given by
{\it secondary models}, which were first put forward by 
Dennison (1980), who suggested that relativistic electrons may be
produced {\it in situ} by inelastic proton-proton collisions through 
production and decay of charged pions. 

There is now increasing evidence that
present radio data, i.e. 
the fine radio properties 
of the radio halos and the halo--occurence,  
may be naturally accounted for by primary 
electron models whereas they may be difficult to be reproduced 
by secondary models (Brunetti, 2003; Kuo et al., 2004).

\begin{itemize}

\item[{\it i)}]
In the framework of {\it primary electron models}, 
merger shocks can accelerate relativistic electrons 
to produce large scale synchrotron radio emission (e.g., Roettiger 
et al., 1999; Sarazin 1999; Takizawa \& Naito, 2000; Blasi 2001).
However, the 
radiative life--time of the emitting electrons diffusing away from these
shocks is so short that they would just be able to produce
relics and not Mpc sized {\it spherical} 
radio halos (e.g., Miniati et al, 2001).
In addition, a number of papers (Gabici \& Blasi 2003; Berrington \& 
Dermer 2003) have recently pointed out that the Mach 
number of the typical shocks produced during major merger events is 
too low to generate non--thermal radiation with the observed fluxes,
spectra and statistics.

\item[{\it ii)}]
Re--acceleration of a population of relic electrons by turbulence 
diffused in the cluster volume and powered 
by major mergers is suitable to explain the very large scale of 
the observed radio emission and is also a promising possibility 
to account for the fine radio structure of the diffuse emission
(Schlickeiser et al.1987; Brunetti et al., 2001a,b).
In this framework, based on relatively simple assumptions,
Ohno, Takizawa and Shibata (2002), and Fujita, Takizawa and Sarazin (2003)
developed specific magneto-turbulent models for Alfv\'enic electron 
acceleration in galaxy clusters.
More recently, Brunetti et al.(2004) presented the first
self-consistent and time-dependent model for the interaction 
of Alfv\'en waves, relativistic electrons, thermal
and relativistic protons in galaxy clusters.
These authors proved that, under some physical conditions,
radio halos and HXR tails may be activated by electron
acceleration due to MHD waves injected during 
cluster merger events.
\end{itemize}

\subsection{Why a statistical model ?}

So far two works have investigated the statistics of the
formation of radio halos from a theoretical point of view.

En\ss lin and R\"ottgering (2002) 
calculated the radio luminosity function of cluster 
radio halos (RHLF).
In a first modelling, they obtained RHLF by combining 
the X-ray cluster luminosity function 
with the radio--halo luminosity -- X--ray luminosity
correlation, assuming that a fraction, $f_{rh}\simeq\frac{1}{3}$, 
of galaxy cluster have radio halos.
Then, in a slightly more accurate modelling, $f_{rh}$ was assumed
to be equal to the fraction of clusters that have recently undergone a strong
mass increase and the radio halo luminosity of a cluster was assumed
to scale with $(1+z)^{-4}$ (due to the incresing IC-losses).

\noindent 
In a more recent paper, Kuo et al. (2004) calculated the 
formation rate and the comoving number density of radio halos in the 
hierarchical clustering scheme. 
The model was based on two morphological criteria to define 
the conditions necessary to the formation of radio halos :
1) the cluster mass must be greater than or equal to a
threshold mass adjusted to observations (Giovannini et al. 1999); 
2) the merger process must be violent enough to distrupt the cluseter core, 
and thus the relative mass increase was required to be 
$\Delta_m\equiv(\Delta M/M)_{th}=0.6$ according to
numerical simulations (Salvador-Sol\`e et al. 1998).
Given the above criteria and making use of the Press-Shechter formalism 
these authors found that a duration of the radio halo phenomenon 
of the order of 1 Gyr would result to be in good agreement with the 
observed probability of formation of radio halos with the mass of the 
parent clusters.

As already pointed out, all these approaches
are based on assumptions in defining the conditions
of formations of radio halos based on observational
correlations and/or mass thresholds.
On the other hand, no effort has been done so
far to model the formation of radio halos and HXR tails
in a self--consistent approach, 
i.e. an approach 
which should model, 
at the same time,
the evolution of the thermal properties of the ICM
of the host galaxy clusters and the generation and evolution
of the non--thermal phenomena.

As mentioned above, one of the ideas that 
is producing the most promising results for
the interpretation of non--thermal phenomena in galaxy
clusters consists in the magneto-turbulent
re--acceleration of relic relativistic electrons
leftover of the past activity occurred within the ICM.

The aim of this paper is thus to obtain a basic modelling 
of the statistical properties of radio halos and HXR tails 
in the above self--consistent approach in the framework 
of the magneto-turbulent re--acceleration scenario.

In order to have a straightforward comparison with published
observational constraints, an Einstein de Sitter (EdS) model 
($H_o=50$ km s$^{-1}$Mpc$^{-1}$, $q_o=0.5$) is assumed in the paper.
In the Appendix A the results are also compared with those obtained in  
a $\Lambda$-CDM model.

\section{The statistical magneto-turbulent Model: Outline}

In this Section we outline the formalism and 
procedures used to develope our statistical model.
The major steps can be sketched as follows :

\begin{itemize}

\item[{\it i)}] {\it Cluster formation}: 
The evolution and formation of galaxy clusters is computed
making use of a semi--analytic procedure based on the hierarchical
Press \& Schechter 1974 (PS) theory of cluster
formation.
Given a present day mass and temperature of the parent clusters, 
the cosmological evolution (back in time) of the cluster
properties (merger trees) are obtained making use of Monte Carlo
simulations. 
A suitable large number of trees
allows us to describe the statistical cosmological
evolution of galaxy clusters.

\item[{\it ii)}] {\it Turbulence in Galaxy Clusters}:
The turbulence in galaxy clusters is supposed to be
injected during cluster mergers. 
 
The energetics of the turbulence injected in the
ICM is calibrated with the $PdV$ work done by the infalling 
subclusters in passing through the volume of the 
most massive one; it basically depends on
the density of the ICM and on the velocity between
the two colliding subclusters.
The sweaped volume in which turbulence is injected
is estimated from the {\it Ram Pressure Stripping} 
(e.g., Sarazin 2002 and ref. therein).
We assume that a relatively large fraction of the turbulence 
developed during these mergers is in the 
form of {\it fast magneto--acoustic waves} (MS waves).
We use these waves since their damping
rate and time evolution basically depend 
on the properties of the thermal plasma which are provided 
by our merger trees for each simulated cluster.

The spectrum of the MS waves depends on many unknown quantities
thus we adopt two extreme scenarios : the first one
assumes a broad band injection of MHD waves (Sect.~4.2),
the second one assumes that turbulence is injected 
at a single scale (Appendix B).
In both cases the spectrum of MS waves 
is calculated solving a turbulent-diffusion equation in the wavenumber 
assuming that the turbulence, injected in the cluster volume for each 
merger event, is injected for- and thus dissipated in a dynamical 
crossing time.

\item[{\it iii)}] {\it Particle Acceleration}: 
We focus on the electron component only because
the major damping of MS waves (which determines the
spectrum of these waves and thus the efficiency of the
particle acceleration, Sect.~5.2) is due to thermal electrons
and thus hadrons cannot significantly affect the
electron--acceleration process 
\footnote{This is different from the case of 
Alfv\'en waves whose damping may be indeed dominated
by the presence of relativistic hadrons (Brunetti et al.2004).}
(see Sect.5.2.2).
We assume a continuous injection of relativistic electrons
in the ICM due to AGNs and/or Galactic Winds.
At each time step, 
given the spectrum of MS waves and 
the physical conditions in the ICM, we
compute the time evolution of relativistic electrons
by solving a Fokker-Planck equation 
including the effect of electron acceleration due to
the coupling between MS waves and particles, and the
relevant energy losses.

\end{itemize}

Given a population of galaxy clusters 
by combining {\it i)-iii)} we are thus able to follow in a
statistical way the cosmological 
evolution of the spectrum of the relativistic electrons 
in the volume of these clusters
and the properties of the thermal ICM.

\section{Cluster formation: 
PS formalism and a Monte Carlo approach}

\subsection{PS formalism and Merger rate}

The PS theory assumes that galaxy clusters form
hierarchically via mergers of subclusters which
develope by gravitational instability of initially small amplitude Gaussian
density fluctuations generated in the early Universe.
Making use of the PS formalism, it is possible to
obtain the probability that a ``parent''
cluster of mass $M_1$ at a time $t_1$ had a progenitor of mass in the range
$M_2 \rightarrow M_2 + d M_2$ at
some earlier time $t_2$, with $M_1 > M_2$ and $t_1 > t_2$.
This is given by (e.g., Lacey \& Cole 1993, Randall, Sarazin \& 
Ricker 2002):

\begin{eqnarray} \label{eq:eps_prob}
{\cal P}(M_2, t_2|M_1, t_1)dM_1 =
\frac{1}{\sqrt{2\pi}}
\frac{M_1}{M_2}
\frac{\delta_{c2}-\delta_{c1}}{(\sigma_{2}^2-\sigma_{1}^2)^{3/2}}
\left| \frac{d\sigma_{2}^2}{dM_2} \right|
\times \nonumber\\
\exp
\left[-\frac{(\delta_{c2}-\delta_{c1})^2}{2(\sigma_{2}^2-
\sigma_{1}^2)}
\right] dM_2 \, ,
\label{mergermerger}
\end{eqnarray}

\noindent where $\delta_{c}(z)$ is the critical linear overdensity 
for a region to collapse at a redshift $z$; for the EdS model it 
is given by:

\begin{equation}  
\delta_{c}(t)=\frac{3(12\pi)^{2/3}}{20}\big(\frac{t_{0}}{t}\big)^{2/3}
\label{delc}
\end{equation}
 
\noindent with $t_{0}$ the present time, 
$\sigma(M)$ is the rms density fluctuation within a sphere of
mean mass $M$. In Eq.~(\ref{mergermerger}) it is 
$\delta_{c1} \equiv \delta_c (t=t_1)$ and
$\sigma_{1} \equiv \sigma ( M_1 )$, with
similar definitions for $\delta_{c2}$ and $\sigma_{2}$.
The standard deviation of matter density fluctuations at the smoothing
scale R ($\sigma(M)$) is given by (e.g., Peebles, 1980):

\begin{equation}  
\sigma(M)=\frac{1}{(2\pi)^3}\int P_{f}(k)\hat{W}^{2}(kR_M)d^{3}k
\label{sigmateo}
\end{equation}

\noindent where $\hat{W}^{2}(kR_M)$ is the Fourier transform of the window 
function, the scale $R_M$ is chosen to contain a mass
$M$, and $P(k)$ is the matter power spectrum at a 
given redshift that can be expressed as:

\begin{equation}  
P_{f}(k)=P_{i}(k)T^{2}(k;z_f)\bigg[\frac{D(z_f)}{D(z_i)}\bigg]^{2}\propto
k^{n}T^{2}(k;z_f)\bigg[\frac{D(z_f)}{D(z_i)}\bigg]^{2}
\label{Pk}
\end{equation}

\noindent where $n$ is the spectral index of the primordial power spectrum
$P_{i}(k)\propto k^{n}$; $T(k)$ is the `transfer function' which
transfer the initial perturbation to the present epoch and $D(z)$ is the 
growth factor of the linear perturbation.
Over the range scales of our interest it is sufficient to consider
a power-law spectrum of the density perturbation given by (Randall, Sarazin \& 
Ricker 2002):

\begin{equation}  \label{eq:sigma}
\sigma (M) = \sigma_{8} \, \left( \frac{M}{M_8} \right) ^{-\alpha} \, ,
\label{sigmam}
\end{equation}

\noindent where $\sigma_{8}$ is the present epoch rms density 
fluctuation on a scale of 8 $h^{-1}$ Mpc, and 
$M_8 = ( 4 \pi / 3 ) ( 8 \, h^{-1} \, {\rm Mpc} )^3 \bar{\rho}$
is the mass contained in a sphere of radius 8 $h^{-1}$ Mpc
($\overline{\rho}$ is the present epoch mean density of the Universe).
The exponent in Eq.(\ref{sigmam}) is given by 
(Bahcall \& Fan 1998) $\alpha = (n+3)/6$, 
$n$ being the slope of the power spectrum of the fluctuations.
Following Randall, Sarazin \& Ricker (2002), it is
$\sigma_8=0.514$ for the EdS models.
\footnote{
By comparing Eq.~\ref{sigmam} with numerical
solutions and analytical fits of Eq.~\ref{sigmateo} (kindly provided
by G.Tormen) we find that, for the typical masses of the
merging subclumps which dominate the injection of cluster turbulence in the ICM
($> 10^{13}M_{\odot}$, Sect.~4), Eq.~\ref{sigmam} is appropriate within a 
few percent.}

It is convenient (Lacey \& Cole 1993) to 
replace the mass $M$ and time $t$ (or redshift $z$) with the
suitable 
variables $S \equiv \sigma^2 ( M )$ and $x \equiv \delta_c ( t )$.
With these definitions, 
$S$ decreases as the mass $M$ increases, and $x$
decreases with increasing cosmic time $t$.

Let 
$K(\Delta S, \Delta x) \, d \Delta S$ be the probability that
a cluster had a progenitor with a mass corresponding to
a change in $S$ of $\Delta S = \sigma_{2}^2 - \sigma_{1}^2$
in the range $\Delta S \rightarrow \Delta S + d \Delta S$ at an earlier
time corresponding to $\Delta x = \delta_{c2}-\delta_{c1}$.
Then, from Eq.(\ref{mergermerger}) one has :

\begin{equation} 
K(\Delta S, \Delta x) d \Delta S = \frac{1}{\sqrt{2 \pi}}
        \frac{\Delta x}{(\Delta S)^{3/2}}
        \exp \left[- \frac{(\Delta x)^2}{2 \Delta S} \right] d \Delta S
\, .
\label{mergermergerK}
\end{equation}

\begin{figure}
\resizebox{\hsize}{!}{\includegraphics{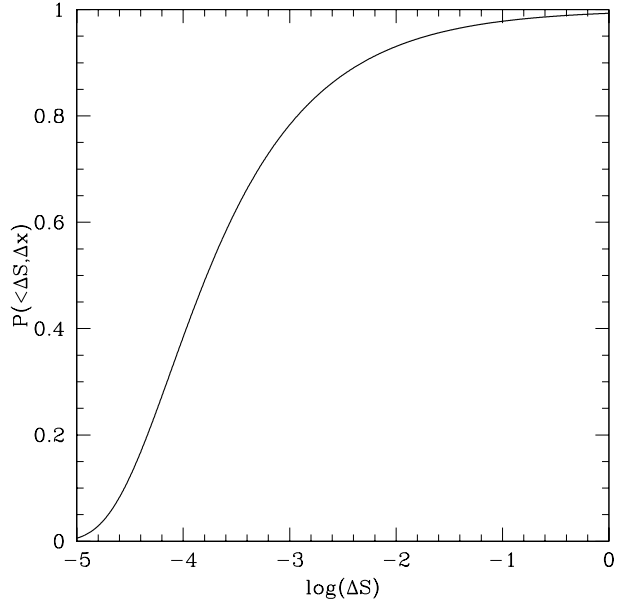}}
\caption[]{Comulative probability distribution 
${\cal P}(<\Delta S,\Delta x)$ as
a function of $log(\Delta S)$.}
\end{figure}

\subsection{Monte Carlo Technique and Merger Trees}

Following a relatively standard procedure adopted 
in the literature
(e.g., Randall et al., 2002; Gabici \& Blasi 2003), 
we employ a Monte Carlo technique to construct merger trees.
Each tree starts at the present time 
with a cluster of mass $M$ and temperature $T$.
We step each simulated cluster back in time, using a small but 
finite time step corresponding to a positive increase $\Delta x$.
The step size determines the value of the minimum mass increment
of the cluster, $\Delta M_c$, which is due essentially to a single 
merger event (Lacey \& Cole 1993) : 

\begin{equation}
(\Delta x)^2 \la \left| {{ d \ln \sigma^2}\over
{d \ln M}} \right| \left( {{\Delta M_c}\over{M}}\right) 
S \, ,
\label{criterion}
\end{equation}

where $M$ is the mass of the cluster at the current time step.
The value $\Delta M_c$ gives the mass of the smallest merging
subcluster we can resolve individually in our trees; 
we choose $\Delta M_c\simeq 10^{12}h^{-1}M_{\odot}$.
Thus mass increments smaller than this value are considered
to be part of the continuous mass accretion
process in galaxy clusters.

In order to follow the probability that a merger with a given
$\Delta S$ (i.e. $\Delta M$) occurs at a given time we make use   
of the cumulative probability distribution of subcluster masses:

\begin{eqnarray}  
{\cal P}( <\Delta S, \Delta x ) =
\int_{0}^{\Delta S} K (\Delta S^{\prime}, \Delta x)
\, d\Delta S^{\prime}
= 
\nonumber\\
{\rm erfc} \left( \frac{\Delta x}{\sqrt{2 \Delta S}} \right) \, ,
\label{mergermergerC}
\end{eqnarray}

where ${\rm erfc} (  )$ is the complementary error function.
The cumulative probability distribution (Fig.~1)
is defined such that
${\cal P}( <\Delta S, \Delta x ) \rightarrow 1$ for
$\Delta S \rightarrow \infty$.

The Monte Carlo procedure selects 
a uniformly-distributed random number, $r$, in
the range 0--1, then it 
determines the corresponding value of $\Delta S$ solving numerically
the equation ${\cal P}( < \Delta S, \Delta x ) = r$
(Fig.~1).
The value of $S_2$ of the progenitor is given by
$S_2 = S_1 + \Delta S$.
The mass of one of the subclusters is given by $\sigma^2 ( M_2 ) = S_2$,
where $\sigma ( M_2 )$ is given by Eq.(\ref{sigmam}),
wheras the mass of the other subcluster is
$\Delta M = M_1-M_2$.

We define $M_{min} \equiv \min(M_2,\Delta M)$ and $M_{max} \equiv
\max(M_2,\Delta M)$. In order to speed up 
the computational procedures, without 
significantly affecting the results, we consider two cases :

\begin{itemize}

\item{i)}
if $M_{min}<1\times10^{13}M_{\odot}$ and $M_{min} < \Delta M_c$,
then the event cannot be identified with a merging subcluster and 
is considered as accretion.
If $M_{min}<1\times10^{13}M_{\odot}$ and $M_{min} > \Delta M_c$,
then the event is considered a very minor merger and 
its contribution to the
injection of cluster turbulence (Sec.~4) is neglected
\footnote{Note that we are interested in describing mergers
of typically $>5 \times 10^{14}$M$_{\odot}$, Sect.~7.}.
In both cases, the mass of the parent cluster is simply 
reduced to $M_2=M_1-M_{min}$ and the next time-step in the 
merger tree starts from $M_2$.

\item{ii)}
if $M_{min}>1\times10^{13}M_{\odot}$ then
the event is treated as a merger 
and we calculate all the physical quantities useful for
the computation of the energy of the turbulence generated during
this event (Sect.4).
In this case, if $M_{min}$ is also greater than a given value of interest
(Sect.7)
we follow back in time the evolution of both the subclusters
(i.e., $M_{min}$ and $M_{max}$) constructing the merger tree for
each subcluster.

\end{itemize}

This procedure is thus iterated until either the mass of the
larger cluster drops below $\Delta M_c$ or 
a maximum redshift of interest $z_{max}$ is reached.
An example of a merger tree obtained from our procedure
(tracing the evolution of the $M_{max}$ clusters only)  
is shown in Fig.~2 as a function of both look back time and
redshift.

Our procedure is basically a {\it Binary Merger Tree Method} 
which does not allow to describe multiple nearly simultaneous
mergers. This simple procedure, however, is sufficient for
our purposes since multiple mergers mainly affect the evolution
of low mass halos at relatively high redshift which are not
interesting for the study of the non-thermal phenomena.
The implementation of more complicated 
{\it N-Branch Tree Methods} can be found in 
Somerville \& Kollatt (1999).

\begin{figure}
\resizebox{\hsize}{!}{\includegraphics{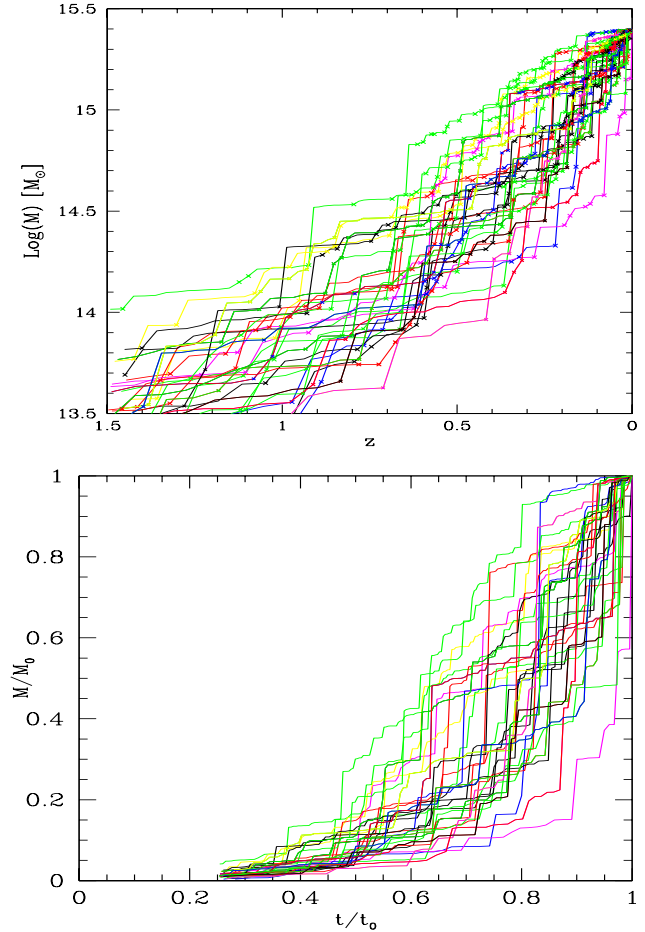}}
\caption[]{Example of Merger Trees obtained from Monte Carlo simulation 
in a EdS universe for clusters with present day mass $M_{0}=2.5\times 10^{15}
M_{\odot}$: a) $Log(M)-z$; b) $M/M_{0}-t/t_{0}$ with t cosmic time; 
$t_{0}$ present time.}
\label{trees}
\end{figure}

\section{Ram Pressure Stripping, turbulence and MHD waves}

\subsection{Turbulence injection rate}

The passage of the infalling subhalos through the
main cluster during mergers induce large--scale bulk 
flows with velocities of the order of $\sim 1000$ km s$^{-1}$ or larger.
Numerical simulations of merging clusters
(e.g., Roettiger, Loken, Burns 1997; 
Ricker \& Sarazin 2001; Tormen et al. 2004)
provide a detailed description of the gasdynamics during
a merger event.
It has been found that subclusters generate laminar bulk 
flows through the sweeped volume of the main clusters which inject 
eddies via Kelvin--Helmholtz instabilities at the interface of the 
bulk flows and the primary cluster gas.
Finally these eddies redistribute the energy of the merger through the
cluster volume in a few Gyrs by injecting random and turbulent
velocity fields.

The impact velocity between the subclusters increases
at the beginning of the merger and then it saturates when
the subclusters interpenetrate each other.
Depending on the initial conditions and on the mass ratio of
the two subclusters, during the merging process the infalling 
halos may be efficiently stripped due to the ram--pressure.
However, the numerical simulations show that
the efficiency of the ram pressure stripping 
is reduced by the formation of a bow shock on the leading age
of the subcluster.
This bow shock forms an oblique boundary layer which slows the
gas flow and redirect it around the core of the subcluster
so that, at least in the case of mergers with mass ratios $<10$,
a significant amount of the subcluster gas is found to be still 
self--bounded after the first passage through the central regions 
of the main cluster (Roettiger et al. 1997; Tormen et al. 2004).

Due to the complicated physics involved in these events, 
the details of the injection and evolution of turbulent motions
in galaxy clusters during merging processes are essentially 
a still unexplored issue. However, turbulence should be basically
driven by the $PdV$ work done by the infalling halos through the
volume of the primary cluster and the turbulent motions should
be initially injected within the volume sweeped by the passage of the
subhalos (e.g., Fujita, Takizawa, Sarazin 2003).
Following this simple scenario, in this Section we estimate the 
rate of turbulence injected during a merger event.
As a necessary approximation (due to the PS formalism) 
in the calculations, we assume that subclusters
undergo only central collisions.

\begin{figure*}
\resizebox{\hsize}{!}{\includegraphics{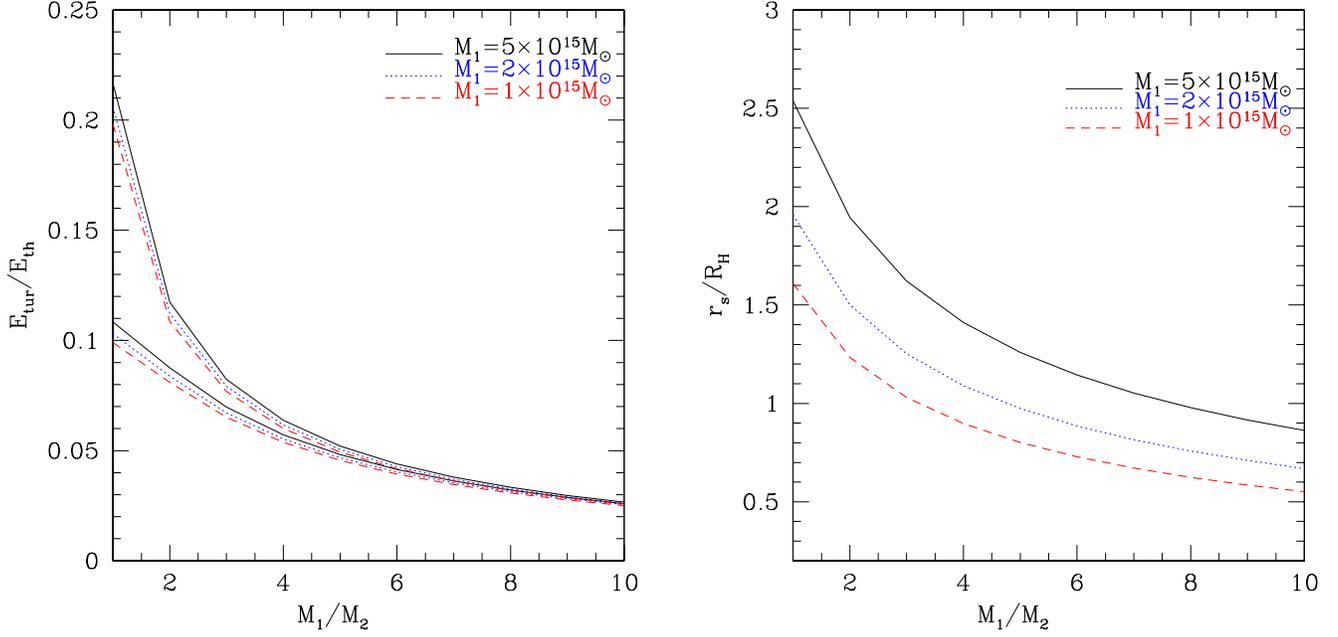}}
\caption[]{
{\bf Panel a)}: Ratio between the energy injected in form of
turbulence and the thermal energy of the system
as a function of the mass ratio of the two subclusters.
Thermal energy is calculated for both the sum of the two
subclusters (lower lines) and for the main cluster
alone (upper lines).
We stress that in the case of a mass ratio $\leq 1.5$ our
approach is quite inadequate because essentially no primary cluster
exists, however these mergers are very rare events and
do not dominate the injection of turbulence in our model.

{\bf Panel b)}: Ratio between the stripping radius and
the radius of the radio halos
(assuming $R_H = 500 h_{50}^{-1}$ kpc)
reported as a function of the mass ratio of the two
subclusters.

In both panels calculations are obtained following the recipes
given in Sect.~4.1 for a $M=5$, (solid lines),
2 (dotted lines), and $1 \times 10^{15} M_{\odot}$
(dashed lines) clusters.}
\label{enrgy_ratio_ok}
\end{figure*}

The relative impact velocity of two subclusters with mass
$M_{max}$ and $M_{min}$ which collide (at a distance $R_{max}$ 
between the centers) starting from an initial distance 
$d_o$ with zero velocity is given by (e.g., Sarazin 2002):

\begin{equation}
v_i\simeq \Bigg(2G\frac{(M_{max}+M_{min})}{R_{max}}
\bigg(1-\frac{1}{\eta_{v}}\bigg)\Bigg)^{1/2}
\label{vi}
\end{equation}

where $d_o=\eta_v R_{max}$, 
$\eta_v\simeq 4\big(\frac{ M_{max}+M_{min}}{M_{max}}\big)^{1/3}$, 
and $R_{max}$ is the virial radius of the main cluster, i.e. :

\begin{equation}
R_{max}=\Bigg[\frac{3M_{max}}{4\pi\Delta_{c}(z)\rho_{crit}(z)}
\Bigg]^{1/3}
\label{Rv}
\end{equation}
$\rho_{crit}(z)$ is the critical density at redshift z and 
$\Delta_{c}(z)=18\pi^{2}\simeq 180$ is the ratio of the average 
density of the cluster to the critical density at redshift z.\\

While the smaller subcluster crosses the larger one, 
it is stripped
due to the effect of the ram--pressure.
The stripping is efficient outside a radius $r_s$ 
(stripping radius) at which equipartition between static 
and ram--pressure is established, i.e. :

\begin{equation}
\overline{\rho}_{max} v_i^2=\frac{\rho_{min}(r_s)K_B 
T_{min}}{\mu m_{p}}
\label{rst}
\end{equation}     

\noindent 
where, as an approximation, 
$\overline{\rho}_{max}$ is fixed at the average density of the ICM
of the larger subcluster :
 
\begin{equation}
\overline{\rho}_{max}=\bigg(\frac{M_{max}}{\frac{4}{3}\pi R_{max}^3}\bigg)\times f_{b} 
\, ,
\label{rhoM}
\end{equation}  

\noindent 
with
$f_{b}=0.25\,(\frac{h}{0.5})^{-3/2}$ the observed barion fraction of clusters
(Ettori \& Fabian 1999, Arnaud \& Evrard 1999). 
Eq.(\ref{rst}) is solved numerically at each merger event 
assuming that the density 
profile of the ICM of the smaller cluster, 
$\rho_{min}$, 
is described by a $\beta$-model (Cavaliere \& Fusco-Femiano, 1976):

\begin{equation}
\rho_{min}(r)=\rho_{min}(0)
\bigg[1+(\frac{r}{r_{c}})^2
\bigg]^{-3\beta_{\rm x}/2}
\label{rhodM}
\end{equation}
 
\noindent 
where the normalization is given by :

\begin{equation}
\rho_{min}(0) =
{{f_b M_{min}}\over {4 \pi}}
\Bigg\{
\int_{0}^{R_{v}^{M_{min}}} dr r^2 
\bigg[1+(\frac{r}{r_{c}})^2
\bigg]^{-3\beta_{\rm x}/2}
\Bigg\}^{-1}
\label{norm}
\end{equation}

\noindent
where a core radius $r_c=0.1 \,\, R_{min}$ and 
$\beta_{\rm x}\simeq 0.8$ are assumed.
\noindent The temperature of the smaller cluster, 
$T_{min}$, in Eq.(\ref{rst}) 
is estimated making use of the observed {\it M-T} relationship 
(e.g., Nevalainen et al. 2000).   
As a general remark we stress that the value of the stripping
radius obtained above would give the mean value of
$r_s$ during a merger and it is not the minimum $r_s$. 
In qualitatively agreement with numerical
simulations, this approach yields
$r_s \rightarrow 0$ in the case of mergers with large mass ratios between
the two colliding subclusters.

\noindent The motion of the smaller cluster through the ICM of the 
main one generates fluid turbulence.
Following Fujita et al.(2003) we assume that turbulence 
is initially injected in the sweeped volume, 
$V_t \sim \pi r_s^2 R_{max}$, with a 
maximum turbulence lenght scale of the order of 
$2\times r_s$.
The total energy injected in turbulence during
a merger event is thus $E_t\simeq \overline{\rho}_{max,s}\,v_i^2 V_{t}$,
where $\overline{\rho}_{max,s}$ is the ICM density of the main cluster
averaged on the sweeped cylinder.
We assume that the duration of the injection is of the order 
of a crossing time, $\tau_{cros}\simeq R_{max}/v_i$, 
then the turbulence is dissipated in a relatively short time.

\noindent
The use of the averaged density of the ICM of the primary cluster,
$\overline{\rho}_{max}$, of the initial impact velocity between
the subclusters, $v_i$, and of the density of 
the main cluster averaged on the sweeped cylinder,
$\overline{\rho}_{max,s}$ in the caluclations of the
injected turbulence is a necessary simplification which 
however guarantees a basic estimate of the averaged 
injected turbulence in the ICM and 
which does not depend on essentially unknown details.
For seek of completeness, in Fig.~3a we report the typical 
ratio between turbulent energy injected by a merger event 
and the thermal energy of the system as a function of the 
mass ratio between the two colliding subclusters; it is found 
that major mergers may channel about 10-15 \% of the thermal energy 
in the form of large scale turbulence.
In Fig.~3b we also report the value of the 
stripping radius as a function of the mass ratio of the
two colliding subclusters. It is found that $r_s$ 
(i.e., the mean value of $r_s$ during a merger event) is 
typically larger than the radius of the 
radio halos, $R_H$, for the merger events
which mainly contribute to the injection of cluster turbulence
in our model.
If the sweeped volume is smaller than that of the radio halo,
we assume that the injected turbulence is diffused over the volume 
of the radio halo, $V_H=\frac{4}{3}\pi R_H^3$, which is basically 
equivalent to assume that the integral cross section of the ensemble 
of minor mergers which occur in a time interval of $\sim$Gyr is 
comparable to $R_H$.

Under these hypothesis, the injection rate per unit volume
of turbulence is given by :

\begin{equation}
\frac{E_{t}}{\tau_{cros}\times
V_H}
\simeq
\frac{\overline{\rho}_{max,s}}{R_{max}}v_i^{3}\bigg(\frac{V_t}{V_H}\bigg)
\label{rate}
\end{equation}

\begin{figure}
\resizebox{\hsize}{!}{\includegraphics{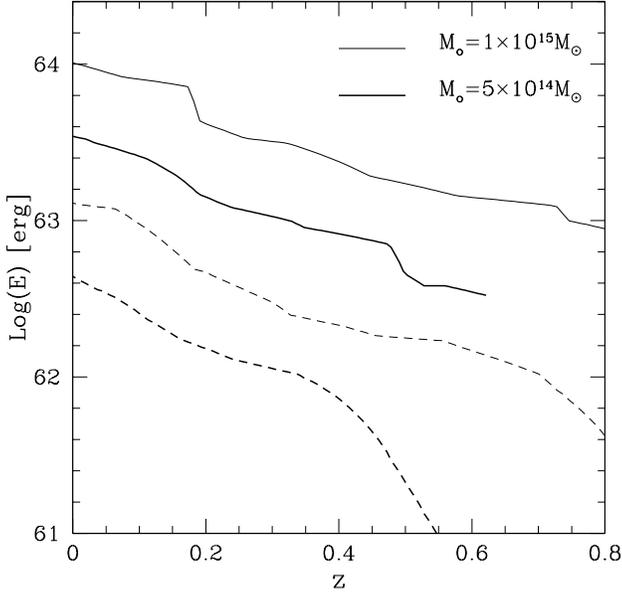}}
\caption[]{Evolution of the thermal energy (solid lines) and of the 
energy injected in fluid turbulence (dashed lines)
integrated during cluster life (at redshift $z$) in typical galaxy clusters.
The thin lines are for a cluster with present time $M_{0}=10^{15}M_{\odot}$ 
and the thick lines are for a cluster with present time $M_{0}=5\times10^{14}
M_{\odot}$.}
\label{enrgy}
\end{figure}

As a relevant example, in Fig.(\ref{enrgy}) we report
the cosmological 
evolution of the thermal energy of galaxy clusters with different masses,
together with the total energy injected in form of 
turbulence in the ICM. The energy in turbulence is calculated by 
integrating the contributions from all the merger events.
The thermal energy of the considered clusters, calculated assuming 
the observed {\it M-T} relation (e.g., Nevalainen et al. 2000)
, increases from about
$10^{62}$erg at $z \sim 1$ to a few $10^{64}$erg at the present 
epoch depending on the mass of the cluster.
As it should be, we note that  
the energy budget injected in turbulence during cluster 
formation is well below the
thermal energy; this indicates the consistency of our calculations.
In particular the turbulent energy is found to be 
$\sim 15\%$ that of the thermal energy in agreement with
recent numerical simulations (Sunyaev et al. 2003) and 
with very recent observational claims (Schuecker et al. 2004).
Finally, as reasonably expected, the energy injected in turbulence 
calculated with our approach is found to roughly
scale with the thermal energy of the clusters.

\subsection{Spectrum of the magnetosonic waves}

\noindent
Cluster mergers should generate magnetosonic (MS) waves
in the ICM, the injected energy and spectrum of these waves depends 
on many unknown quantities.
A reasonable attempt is to assume that a fraction, 
$\eta_t$, of the energy of the turbulence (Sect.~4.1)
is in the form of MS waves.
We shall consider two extreme scenarios:

\begin{itemize}

\item[{\it i)}] in the first one we assume that
MS waves are driven
by the plasma instabilities (e.g., Eilek 1979, and 
ref. therein) which develope in the turbulent 
field generated during cluster mergers. In this case
MS waves may be injected over a broad range of scales.
Here, we shall adopt a simple power law injection spectrum 
of these waves: $I(k)=I_o k^{-a}$ for $k> k_{min}\sim \pi/r_s$;

\item[{\it ii)}]
in the second one, we assume that 
MS waves are basically injected at a 
single scale, $k=k_{min}\sim \pi/r_s$, from which a MHD turbulence 
cascade is originated.

\end{itemize}

In both cases the decay time of the MHD turbulence at the 
maximum/injection scale, $L_{inj}\sim 2 r_s$, can be estimated as 
(e.g., Appendix B) 
$\tau_{kk}(L_{inj}) \sim {{ r_s }\over{\eta_t v_i}}$,
one has :

\begin{equation}
\tau_{kk}({\rm Gyr}) \sim 1 \times 
( {{v_i}\over{2\cdot 10^3 {\rm km/s}}})^{-1} 
({{r_s}\over{500 {\rm kpc}}}) 
({{\eta_t}\over{0.25}})^{-1}
\label{tempo}
\end{equation}

\noindent
which is of the order of a crossing time and thus
allows the MHD turbulence to diffuse filling a volume of the
order of that of radio halos (or larger) with a fairly
uniform intensity.

In the following we focus on the first scenario, while in the
Appendix B we consider the second picture.
Appendix B demonstrates that these two extreme scenarios lead to
very similar results and thus that, in our model, the details of 
the injection process of the MS waves do not appreciably
change the conclusions. 

In the case in which a power law spectrum of MS waves is injected
in the ICM, one has :

\begin{equation}
\int_{k_{min}}^{k_{max}} I_o k^{-a}dk=
\eta_t \frac{E_{t}}{\tau_{cros}\times V_H}
\label{Ioint}
\end{equation}

\noindent with $k_{min} \sim \pi/r_s$ and $k_{max}=\Omega_p/v_M$
$\Omega_p$ being the proton cyclotron frequency and $v_M$
the magnetosonic velocity (Eq.~\ref{dispersionms}).
From Eq.~(\ref{Ioint}) we find:

\begin{eqnarray}
I_o=\left\{\begin{array}{ll}
\frac{E_{MS}}{\tau_{cros}}\times\frac{(a-1)}{V_{H}}k_{min}^{-(1-a)}&
( r_s\leq R_H)\\
& \\
\frac{E_{MS}}{\tau_{cros}}\times\frac{(a-1)}{V_{H}}k_{min}^{-(1-a)}\times
(\frac{R_H}{r_s})^{2} & (r_s>R_H)\\
\end{array}\right.
\label{Io}
\end{eqnarray}
 
Thus, the injection of the MS waves is obtained by combining
Eqs.(\ref{rate}) and (\ref{Io}).
$\eta_t$ is the first free parameter of our model, in order to
have a self--consistent modelling it should be $\eta_t <1$.
  
In general, the spectrum of MHD waves injected in the ICM evolves 
due to wave--wave and wave--particle coupling. 
The combination of these processes produces a modified
spectrum of the waves, $W_{\rm k}(t)$.
In the quasi linear regime the spectrum of the waves can 
be calculated solving a continuity equation in the wavenumber space:

\begin{eqnarray}
{{\partial W_{\rm k}(t)}\over
{\partial t}} =
{{\partial}\over{\partial k}}
\left( D_{\rm kk} {{\partial W_{\rm k}(t)}\over{\partial k}}
\right)
-\sum_{i=1}^n \Gamma^{\rm i}(k) W_{\rm k}(t) 
\nonumber\\
+
I_{\rm k}(t)
\label{turbulencems}
\end{eqnarray}

\noindent
The first term on the right hand describes the wave--wave interaction, 
with diffusion coefficient $D_{\rm kk} = k^2/\tau_s$ (with $\tau_s$ the 
spectral energy transfer time).
The second term in Eq.(\ref{turbulencems})
describes the damping with the relativistic and thermal particles 
in the ICM, and the damping due to thermal viscosity.
In the following, we shall neglect the term due to 
the wave--wave interaction, this is justified provided that
the time--scale of the dampings are smaller than that of the 
wave--wave cascade (or comparable), at least for the range
of scales which contribute to the
acceleration process (assuming spectra flatter than
$W_k \propto k^{-2}$; see also Appendix B).
Under physical conditions typical of the ICM the most
important damping in the collisionless regime
is that with the thermal electrons (e.g., Eilek 1979; see also
Sect.~5.2.2).
An estimate of this damping rate (for $v_A < v_M$ as in the ICM) 
is given by Eilek 1979; a relatively simple formula, 
consistent whitin a 10\% with the Eilek's results, is:

\begin{eqnarray}
\Gamma_{th,e}=
\sqrt{32\pi^{3}}\, n_{th}\,(m_e K_B T)^{1/2}
\big(\frac{v_M}{B}\big)^{2}
\frac{{W}_k^B}{{W}_k}\,{\cal I}(x)\, k
\label{dampthms}
\end{eqnarray}

\noindent
where $n_{th}$ is the number density of the thermal electrons, 
${W}^{B}_k$ is the 
turbulent magnetic energy density, ${W}^{B}_k={W}_k\times
(1+8\pi P/B^2)^{-1}$, with $P\simeq 2 n_{th}K_B T$ the thermal pressure, 
$B$ the plasma magnetic field (e.g., Barnes \& Scargle 1973) and 
${\cal I}(x)$ is a numerical value given by : 

\begin{equation}
{\cal I}(x)=2\,\int_{1}^{+\infty}dx \,\Big(\frac{1}{x}-\frac{1}{x^3}\Big)
\, e^{-[x^2(\frac{v_M}{v_{th}})^2]}
\label{Int}
\end{equation}

where $x=\frac{p_{\Vert}}{m_e v_M}$, with $p_{\Vert}$ the component 
of the momentum of the thermal electrons along the magnetic field lines
and $v_{th}=(2\,K_B\,T/m_e)^{1/2}$.

Since for each merger event we are interested in the evolution 
of the spectrum of the injected waves on a time scale of 
$\sim 1$ Gyr, which is orders of magnitude 
longer than the typical time scales of the damping processes, 
the spectrum of the waves is expected to approach a 
stationary solution ($\partial W_{\rm k}/ \partial t =0$).
From Eq.~(\ref{turbulencems}) this solution is given
by :

\begin{eqnarray}
W_{\rm k}
\simeq
\frac{I(\rm k)}
{\Gamma_{\rm th,e}(k)}
=\frac{I(\rm k)}{f(T)\,k}
\label{ms-stationary}
\end{eqnarray}

\noindent
In Sect.~5.2 we will derive the efficiency of electron acceleration
due to the MS waves.
Here we would just point out that the acceleration time, 
$\tau_{\rm acc}$, depends on :

\begin{equation}
\tau_{\rm acc}^{-1} \propto
\int_{k_{min}}^{k_{max}}
k W_{\rm k} dk
\label{tauaccpro}
\end{equation}

which leads (making use of Eqs.~\ref{Io} and \ref{ms-stationary})  
to the nice result that the acceleration time in our model,
and under our assumptions, does not depend on the slope 
of the injection spectrum of MS waves (which depends on 
basically--unknown details of the injection mechanism) 
and on the value of $k_{min}$.

\subsection{Spectrum of MS waves during cluster formation}

In this Section we estimate the spectrum of MS waves resulting
from the combination of the contributions of several mergers during
the process of cluster formation.
For a given galaxy cluster, we define $z_{i}^j$ to be
the redshift at 
which the $j^{th}$ merger event starts.
For an Einstein-De Sitter model, the corresponding time, $t_i^j$,
is :

\begin{equation}
t_i^j=\frac{2}{3 H_o}\frac{1}{(1+z_i^j)^{3/2}}
\label{ti}
\end{equation}

\noindent 
In our simple modelling we assume that the duration time of a merger
is of the order of a crossing time, $\Delta t=t_{cros}^j$,
and that turbulence is injected during this time interval and then suddenly 
dissipated via damping processes
\footnote{Note that if the injection time is slightly longer 
than $t_{cros}$ then the probability to combine 
the effect of several mergers increases and the efficiency
of the model would slightly increase (see above, Eq.\ref{Wtot}).}.
Thus the turbulence injected during
the $j^{th}$ merger is dissipated at time $t_f^j = t_i^j + t_{cross}^j$
and the corresponding redshift is given by :

\begin{equation}
z_f^j=(\frac{2}{3 H_o t_f^j})^{2/3}-1
\label{zf}
\end{equation}

\noindent 
We describe the spectrum of the MS--waves
established during the $j^{th}$
merger event as :

\begin{equation}
W_k^j(z)=W_k^j(z_i^j)\times S^j(z)
\label{Wj}
\end{equation}

\noindent were $S^j(z)$ is a step function defined as :

\begin{eqnarray}
S^j(z)=\left\{\begin{array}{ll}
1 & ( z_f^j< z <z_i^j)\\
 & \\
 0 & ( {\rm otherwise} )\\
\end{array}\right.
\label{contributionj}
\end{eqnarray}

According to the hierarchical scenario adopted in this paper,
clusters undergo several merger events which contribute 
to the injection of turbulence yielding a combined spectrum
of MS--waves.
Since under stationary conditions and neglecting the 
wave--wave interaction term, Eq.(\ref{turbulencems}) is a
linear differential equation, 
the spectrum of the MS--waves resulting from the combination of 
the different merger events is given by the sum of all
the contributions (Eq.\ref{Wj}), i.e. :

\begin{equation}
{\cal W}_{\rm k}(z)=\sum_{j} W_k^j(z_i^j)\times S^j(z)
\label{Wtot}
\end{equation}

\section{Particle Evolution and Acceleration}

V\"olk et al. (1996) and 
Berezinsky et al. (1997) have shown that cosmic 
ray protons are most likely accumulated in galaxy 
clusters as their diffusion time scale is
of the order or larger than the Hubble time.
An even stronger conclusion 
can be applied to the case of the
relativistic electrons injected in the ICM.
Indeed, by assuming a Kolmogorov spectrum for the turbulent field 
the parallel diffusion length, $L_{D}$, of relativistic electrons 
is given by (e.g., Brunetti 2003):

\begin{equation}
L_{D} \simeq 
120 \,
\left(
{{ \gamma/10^2 }\over {B_{\mu G}}}
\right)^{ {{1}\over{6}} }
\left(
{{l_o }\over { 100\, {\rm kpc} }}
\right)^{ {{3}\over{4}} }
{{ \tau_D }\over{ {\rm Gyr} }}
\,\,\,({\rm kpc})
\label{lparkol}
\end{equation}

where $l_o$ is the maximum coherence scale of the 
magnetic field fluctuations.
We notice that $\gamma \sim 100-300$ relic relativistic 
electrons cannot diffuse more than 200 kpc during their life--time 
(a few Gyrs, e.g., Sarazin 1999).
When turbulence is generated in galaxy clusters
the ensuing increase of the particle-wave 
scattering frequency would further reduce 
the net diffusion coefficient.
Only in the case of very strong turbulence anomalous
particle turbulent diffusion may allow electrons to diffuse
over larger scales (e.g., Duffy 1994; En\ss lin 2003), but
the requested conditions are well beyond those assumed
in the present paper.

\noindent
Thus, we can safely assume 
that electrons injected by some mechanism
in the ICM simply follow the thermal plasma and magnetic 
field.
Under this condition, the time evolution of relativistic electrons 
with isotropic momentum distribution is provided
by a Fokker-Planck equation (e.g., Tsytovich 1966; 
Borovsky \& Eilek 1986) for the electron
number density :

\begin{eqnarray}
{{\partial N(p,t)}\over{\partial t}}=
{{\partial }\over{\partial p}}
\left[
N(p,t)\left(
{ \big| {{dp}\over{dt}} \big|}_{\rm rad} + 
{\big| {{dp}\over{dt}} \big| }_{\rm c}
-{2\over{p}} D_{\rm pp}
\right)\right] +
\nonumber\\
{{\partial }\over{\partial p}}
\left[
D_{\rm pp}
{{\partial N(p,t)}\over{\partial p}}
\right] +
Q_e(p,t)
\label{elettroni}
\end{eqnarray}

where $D_{\rm pp}$ is the electron 
diffusion coefficient in the momentum space 
due to the interaction with the MS waves, 
${dp/dt}_{\rm i}$ and ${dp/dt}_{\rm rad}$ are the terms due to 
ionization and radiative losses 
and $Q_e$ is an isotropic electron source term.

\subsection{Particle injection}

A relevant contribution to the injection of cosmic rays in clusters of 
galaxies comes from Active Galactic Nuclei (AGN). AGNs indeed inject in 
the ICM a considerable amount of energy in relativistic particles and also 
in magnetic fields, likely extracted from the accretion power of their central 
black hole (En\ss lin et al., 1997). 

Powerful Galactic Winds (GW) can inject relativistic particles and 
magnetic fields in the ICM (V\"olk \& Atoyan 1999). Although the present day 
level of starburst activity is low and thus this mechanism is not expected to 
produce a significant contribution, it is expected that these winds were more 
powerful during early starburst activity. Some evidence that powerful GW were 
more frequent in the past comes from the observed iron abundance in galaxy 
clusters (V\"olk et al. 1996).

In addition, cluster formation
is also believed to provide a possible contribution to the
injection of cosmic rays in the ICM due to the formation
of shocks which may accelerate relativistic particles
(Blasi 2001; Takizawa \& Naito, 2000; Miniati et al., 2001; 
Fujita \& Sarazin 2001).
The efficiency of this mechanism is related to the Mach number 
of these shocks which is an issue still under debate.
Semi--analytical calculations based on PS-Monte Carlo techniques 
(Gabici \& Blasi 2003; Berrington \& Dermer 2003) find that the
bulk of the shocks have Mach numbers
of order 1.5, as also observed by {\it Chandra}
(e.g., Markevitch et al. 2003).
First cosmological numerical simulations 
found that the Mach number of
merger and internal flow shocks peacks
at ${\cal M} \sim 5$ (Miniati et al. 2000),
and that the bulk of the energy is dissipated at 
$4 \leq {\cal M} \leq 10$ shocks (Miniati 2002).
On the other hand, 
more recent numerical simulations find more weak shocks with 
the bulk of the energy dissipation (thermal energy and
cosmic rays) at internal shocks with 
$2 \leq {\cal M} \leq 4$ (Ryu et al., 2003).
However, the comparison with semi-analytical calculations appears difficult 
because of a different classification of the shocks in the two approaches. 

Independently from the specific scenario adopted for the
injection of relativistic particles, 
a power law spectrum for the injection rate
of relativistic electrons up to a maximum momentum,
$p_{\rm max}$, can be reasonably assumed :

\begin{equation}
Q_{\rm e}(p,t)=K_{\rm e}(t) p^{-s}
\label{qpe}
\end{equation}

We parameterize the injection rate by assuming that the total energy 
injected in cosmic ray electrons (for $p > p_{\rm min}$)
during the cluster life is a fraction, $\eta_e$, of the total thermal energy
of the cluster at $z=0$, i.e. :

\begin{equation}
\eta_e = {{ c }\over{{\cal E}_{\rm th}}}
\int_{t=t(z)}^{t=t(0)}
d\tau \int_{p_{\rm min}}^{p_{\rm max}}
Q_e(p,\tau)\,p\, dp
\label{kene}
\end{equation}

where ${\cal E}_{\rm th}$ is the present day
thermal energy density of the ICM.

The injection rate should depend on the 
number and energetics of AGNs and GWs in galaxy clusters
which are expected to be considerably larger
at high redshifts.
However, since electrons injected at relatively high redshifts
cool very rapidly because of the combination 
of high energy losses and low efficiency of the
particle acceleration mechsnism (Fig.~5), 
only the electrons injected at relatively low redshifts can
be accelerated and therefore contribute to the non-thermal
emission observed at $z<0.2$.
As a simplification, we adopt a constant injection 
rate of electrons so that (for $s>2$) the normalization of the 
spectrum of the injection rate is given by :

\begin{equation}
K_{\rm e} \sim
{{ s-2 }\over{ c}} \eta_e
{\cal E}_{\rm th}
p_{\rm min}^{s-2} \tau_H^{-1}
\label{ke}
\end{equation}

\noindent
where $\tau_H$ is the Hubble time.
$\eta_e$ is the second free parameter in our model.
In the following we use $s=2.5$, $p_{\rm min}/mc=60$,
and $p_{\rm max}/mc=10^4$;
as discussed in Sect.~8 the basic results of our model
do not depend on the values adopted for these parameters.

\subsection{Energy gains and losses of the relativistic particles}

Fast MS waves can accelerate
relativistic particles via particle--wave
interaction.
This interaction occurs with both thermal (expecially) and
relativistic particles via Landau damping, 
the strongest damping of MS waves in
the collisionless regime. The necessary condition  
(Melrose 1968; Eilek 1979) is $\omega - k_{\Vert} v_{\Vert}=0$, 
where $\omega$ is the frequency of the wave, $k_{\Vert}$ is the 
wavenumber projected along the magnetic field, and $v_{\Vert}=v \mu$ 
is the projected-particle velocity.
Under the typical conditions
of the ICM, the dispersion relation for MS waves in an isotropic 
plasma is given by (e.g., Eilek 1979):

\begin{equation}
\frac{\omega^2}{k^2}=
v_M^2\simeq 
\frac{4}{3}v_{ion}^2+v_A^2
\label{dispersionms}
\end{equation}

\noindent
where $v_{ion}$ and $v_A$ are the ion-sound velocity
and the Alfv\'en velocity, 
respectively.
In a isotropic distribution of waves and
particle momenta the 
diffusion coefficient in the momentum space,
for $v_A < v_M$, is given by Eilek (1979).
This can be expressed as:

\begin{equation}
D_{\rm pp}(p,t)\simeq
4.45 \, \pi^{2}\,{{ v_M^2 }\over{c}}
{{p^2}\over{B^2}}
\int_{k_{min}}^{k_{max}}
k {\cal W}^{B}_k(t)
dk
\label{dppms}
\end{equation}

The acceleration time scale, which in this case
does not depend of the particle energy, 
is given by :

\begin{equation}
\tau_{\rm acc}^{-1}
= \chi
\simeq 
2 {{ D_{\rm pp} }\over{
p^{2} }}
\label{chi}
\end{equation}

\noindent
and
thus the systematic energy gain of particles
interacting with MS waves is given by :

\begin{equation}
\left( {{ d p }\over{d t}}\right)_{\rm acc}^{sys}
=
\chi \, p
\label{acc}
\end{equation}

\subsubsection{Electrons}

Relativistic electrons with momentum 
$p_{\rm e} = m_{\rm e} c \gamma$
in the ICM lose energy through ionization losses and Coulomb 
collisions (Sarazin 1999):

\begin{equation}
\left( {{ d p }\over{d t}}\right)_{\rm c} 
=\, -3.3 \times 10^{-29} n_{\rm th}
\left[1+ {{ {\ln}(\gamma/{n_{\rm th}} ) }\over{
75 }} \right]
\label{ion}
\end{equation}
\noindent
where $n_{\rm th}$ is the number density of the thermal plasma.

Relativistic electrons also 
lose energy via synchrotron emission and inverse 
Compton scattering off the CMB photons:

\begin{eqnarray}
\lefteqn{
\left( {{d p}\over{d t}}\right)_{\rm rad}=\, -4.8\times10^{-4} p^2
\left[\left({{ B_{\mu G}}\over{
3.2}}\right)^2{{\sin^2\theta}\over{2/3}}
+(1+z)^4 \right] }{}
\nonumber\\
& & {} \,\,\,\,\,\,\,\,\,\,\,\,\,\,\,\,\,\ 
= -\frac{\beta p^{2}}{m_e\,c}
\label{syn+ic}
\end{eqnarray}
\noindent

where $B_{\mu G}$ is the magnetic field strength in $\mu G$ and $\theta$ 
is the pitch angle of the emitting electrons; in case of efficient 
isotropization of the electron momenta it is possible to average over
all possible pitch angles, so that  $<\sin^2\theta> = 2/3$.

Eqs.(\ref{dppms}), (\ref{ion}) and (\ref{syn+ic}) gives the coefficient
of the Fokker-Plank equation (\ref{elettroni}).

\subsubsection{Protons}

For seek of completeness in this Section we briefly discuss
the case of relativistic protons.

\noindent
The main channel of energy losses for these 
particles is represented
by inelastic proton-proton collisions, which is a threshold reaction that
requires protons with kinetic energy larger than $\sim 300$ MeV.
The time scale of this process is 
$\tau_{pp} \sim 30 (n_{\rm th}/10^{-3})^{-1}$Gyr. 
The interactions with the ICM may generate an appreciable flux of 
gamma rays and neutrinos, in addition to a population of secondary 
electrons (Blasi \& Colafrancesco 1999). 

\noindent
Protons which are more energetic than the thermal electrons
lose energy due to Coulomb interactions.
For trans-relativistic and sub-relativistic protons  
this channel can easily become
the main channel of energy losses in the ICM.

In our calculations we focus on the population of relativistic
electrons, do not consider proton acceleration, 
and neglect the effect of these particles 
on the efficiency of the electron acceleration.

The resonance condition $v_M k = k_{\Vert} v_{\Vert}$
implies that only a very small fraction of MS 
waves (those making an angle $\sim$89--91 degrees with the 
local B-field) cannot be damped by the 
thermal electrons, but only by the relativistic particles 
(protons and electrons), while outside this narrow cone the damping 
due to the thermal electrons should be the strongest 
one (e.g., Eilek 1979).
As a consequence, since in our calculations we assume a continuous pitch
angle isotropization (e.g., Miller et al.~1996)
and an isotropic distribution of MS waves
which propagate in a complex geometry of the field lines,
the damping of MS waves should be dominated by 
the effect due to the thermal electrons in the ICM.

\noindent
As an example, we assume that MS waves are injected in the central 
$\sim$1 Mpc$^3$ of a massive cluster for 0.5--1 Gyr with
a total energy budget of the order of that of
the thermal ICM within the same region.
We calculate particle acceleration and find that about 
$\sim 4-10 \%$ of the energy flux of these waves 
is channelled into the acceleration of relativistic protons 
(assuming an initial energy density of these particles of the
order of few \% of the thermal energy density and 
$N_p(p)\propto p^{-2.2}$): this corresponds to
$\leq 1\%$ of the total thermal energy of the cluster. 
At the same time, we find that only $\sim 0.1 \%$ of the energy 
flux of the MS waves should be channelled into the acceleration 
of relativistic electrons to produce an HXR luminosity of 
$\sim 10^{43}$erg s$^{-1}$ from the same volume 
($\eta_e \sim 0.003$, Fig.~6).
Consequently the $90-95\%$ of the energy flux of the MS waves
is channelled into the thermal electrons and thus  
the resulting spectrum of these waves may be estimated with
good approximation by Eq.(\ref{ms-stationary}).

Detailed time dependent calculations which include 
electron and proton acceleration due to MS waves and a comparison
with the case of Alfv\'en waves will be given in a forthcoming
paper (Brunetti et al., in prep.).

\section{Radio Halos and HXR tails}

\subsection{Cluster evolution and electron spectrum}

In this Section we combine the formalism developed
for the evolution of the turbulence  
(Sects.~3 \& 4) with the recipes for particle acceleration
and evolution (Sect.~5) to model
the cosmological evolution of the spectrum of the 
relativistic electrons in galaxy clusters.

The electron--acceleration coefficient, due to the effect of MS 
waves at redshift $z$, is obtained by combining  Eq.~(\ref{chi}) 
with Eqs.~(\ref{dppms}, \ref{Io},
\ref{ms-stationary}, \ref{Wtot}):

\begin{eqnarray}
\lefteqn{\chi(z)
\simeq {{ 2.23 \times 10^{-16} \eta_t}
\over{ (R_H/500{\rm kpc})^3}}
\sum_{j}
\Bigg[
\Big( 
{ {M_{max} +M_{min} }\over{
2\times 10^{15} M_{\odot}}}
\,\,
{{ 2.6 {\rm Mpc} }\over
{ R_{max}}} 
\Big)^{3/2} }{}
\nonumber\\
& & {} 
\times
\frac{(r_s/500 {\rm kpc})^2}{(kT/7 {\rm keV})^{1/2}} 
\Bigg]_{j}
\times 
\left\{
\begin{array}{ll}
1 & {\rm if}\, r_s \leq R_H \\
 & \\
(R_H/r_s)^2 & {\rm if}\, r_s > R_H \\
\end{array} \right\}_j
\label{Dppt}
\end{eqnarray}

where only mergers which contribute to the turbulence
spectrum at redshift $z$ (Sect.~4.3, Eq.\ref{Wtot}) are
considered.
The evolution of the electron spectrum is thus obtained
from the numerical solution of the 
Fokker-Planck equation (Eq.~\ref{elettroni})
by adopting the values of the coefficient $D_{pp}$ 
(Eq.~\ref{chi} and Eq.~\ref{Dppt}) 
and of the energy loss terms (Eqs.~\ref{ion}--\ref{syn+ic}) 
at each redshift.

\begin{figure}
\resizebox{\hsize}{!}{\includegraphics{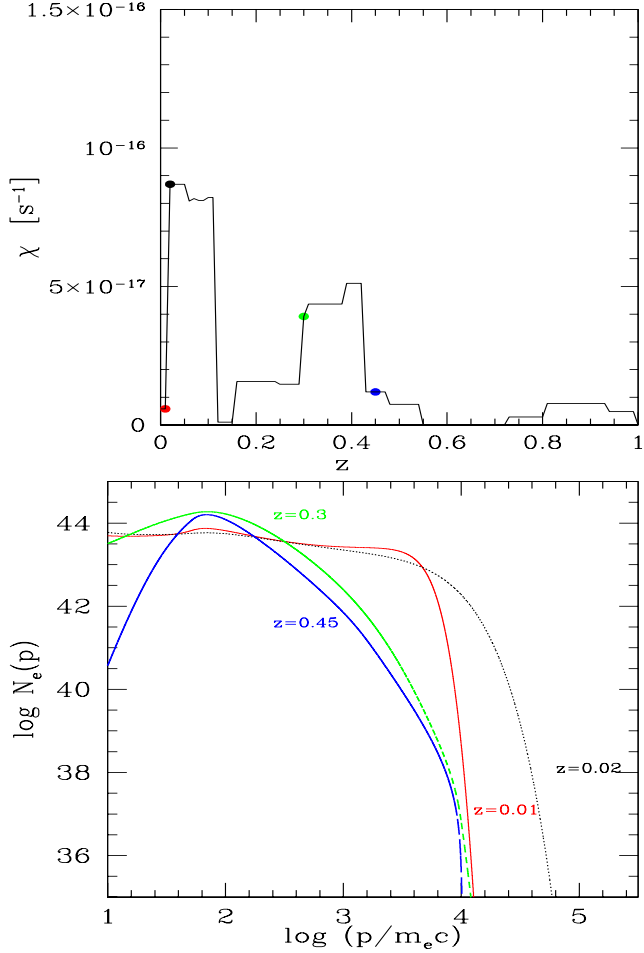}}
\caption[]{{\bf Top panel}: evolution with redshift of 
the electron-acceleration coefficient due to MS waves as obtained 
from Eq.\ref{Dppt} for a cluster of 
$M_{0}=1\times10^{15} M_{\odot}$ at the present time.
{\bf Bottom panel}: electrons spectra (in arbitrary units)
calculated at different redshifts
(also marked in the top panel) for the same cluster.
Calculations are performed for $s=2.5$, $B=0.5 \mu$G,
and $\eta_t=0.26$.}
\label{chispe}
\end{figure}

In Fig.(\ref{chispe}) we report an example of the time
evolution of the electron--acceleration coefficient obtained
for a typical massive cluster (top panel) and the corresponding 
spectra of the electrons at different relevant times (bottom panel):
an increase of the acceleration coefficient produces an increase of 
the maximum energy of the electrons.
The reported results indicate that cluster--merger activity at low redshift 
can generate an increse of the cluster turbulence
which may be sufficient to accelerate electrons up to 
$\gamma >> 10^3$, 
necessary to produce synchrotron radiation in the radio band.
It should be noticed that electrons
are accelerated (and cool) with a delay time (of the order of
the corresponding electron--acceleration time $\sim \chi^{-1}$) 
with respect to the abrupt increases (decreases) of the values of the
acceleration coefficient.

\begin{figure}
\resizebox{\hsize}{!}{\includegraphics{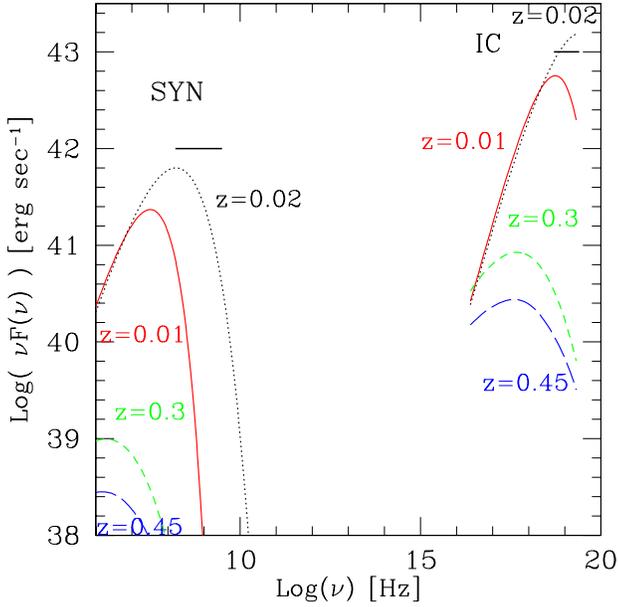}}
\caption[]{Broad band synchrotron (SYN) and Inverse Compton (IC)
spectra calculated for the case reported in Fig.\ref{chispe} 
and for $\eta_e=0.003$, $R_H=500$ kpc, and $B=0.5\mu$G.
Horizontal bars give the radio (used for radio halos)
and HXR observational range of frequencies.}
\label{synic}
\end{figure}

In Fig.(\ref{synic}) we show the broad band 
non--thermal emission (synchrotron and IC)
from the galaxy cluster reported in Fig.~\ref{chispe}
assuming $\eta_e=0.003$ and $B=0.5\mu$ G 
(for the sake of completeness synchrotron and IC equations 
are given in Appendix C).   
The aim of this Figure is to show
that synchrotron (and IC) luminosities of the order 
of those of the most luminous radio halos can be reasonably obtained. 
On the other hand, it should be stressed that
the synchrotron spectrum reported in Fig.(\ref{synic}) 
is obtained assuming a constant value of $B$ through the cluster 
volume. More reasonable calculations should assume 
a radial gradient of the magnetic field strength which  
causes a stretching in frequencies of the synchrotron spectral shape
with respect to that of Fig.(\ref{synic}) 
(e.g., Brunetti et al. 2001; Kuo et al. 2003).

The synchrotron emitted power from radio halos
is also expected to increase with increasing the mass of
the parent clusters.
Indeed the bolometric synchrotron power roughly scales as 
$P_{\rm syn} \propto B^2 \gamma_{\rm b}^2 N_{\rm e} R_H^3$
where $\gamma_b=\frac{\chi}{\beta}$ is the maximum energy 
of the accelerated electrons
($\beta$ is the total energy--loss coefficient,
Eq.~\ref{syn+ic}) and $N_{\rm e}$ is the number of
relativistic electrons in the cluster emitting volume.
During major mergers, from Eq.(\ref{Dppt},
with $T\propto M^{a}$, $a \sim 0.55-0.67$)
one has :

\begin{equation}
P_{\rm syn} \propto
N_{\rm e} M^{2-a} g(r_s,R_H)^2
{ { B^2 }\over{ (B^2 + B_{\rm cmb}^2 )^2 }}
\label{power}
\end{equation}

\noindent
where $g(rs,R_H)$ ($g=r_s^2$ for $r_s \leq R_H$ and 
$g=R_H^2$ for $r_s > R_H$) is a slightly increasing function
of cluster mass.
Thus assuming $B < 3 \mu$G one can find that the synchrotron power
would roughly scale 
as $P_{\rm syn} \propto M^{b(M)} < N_{\rm e} B^2 >_{(M)}$,
where $b(M) \sim 1.5-2.5$ 
($b=1.5$ for $M \geq 3 \cdot 10^{15} M_{\odot}$
and $b=2.5$ for $M < 10^{15} M_{\odot}$),
and $< N_{\rm e} B^2 >_{(M)}$ is expected to increase with
cluster mass.
Although studies of the 
synchrotron power -- mass (or temperature) correlation
for radio halos use the monochromatic synchrotron emission
at 1.4 GHz (e.g., Liang et al. 2000;
Govoni et al. 2001; Giovannini \& Feretti 2002), 
the above expectations seem to be
in line with observations.

\subsection{On the required values of $\eta_t$ and $\eta_e$}

In this Section we derive the range of values of the two free parameters
of our model, $\eta_t$ and $\eta_e$, which provide a reasonable 
agreement with the general properties of radio halos.
In order to check the reliability of the obtained values, 
these are then compared with independent findings
and general expectations from both analytical and numerical
calculations.

The first free parameter is $\eta_t$ which is defined as 
the fraction of the fluid turbulence in MS waves.
The value of $\eta_t$ drives the efficiency of the electron
acceleration and thus the resulting maximum energy of electrons,
$\gamma_b=\frac{\chi}{\beta}$,
($\beta$ is the total energy--loss coefficient,
Eq.~\ref{syn+ic}), 
and the maximum synchrotron emitted frequency 
$\nu_b=\frac{3}{4\pi}
\frac{e B}{m_e c} \gamma_b^2 $.
Under the assumption that the losses of
the electrons are dominated by the IC mechanism,
the acceleration coefficient
is thus related to the break frequency by :

\begin{equation}
\chi \simeq 6.3 \times 10^{-21}\nu_b^{1/2}B_{\mu G}^{-1/2}(1+z)^4
\label{rel4}
\end{equation}

The values of $\eta_t$ are constrained by requiring that
the accelerated electrons can produce synchrotron
radiation in the radio band with the spectral shape 
observed in the case of radio halos, i.e. with spectral
index $\alpha= 1.1-1.5$ between 327 and 1400 MHz (e.g., Kempner \&
Sarazin 2001).
The synchrotron spectral index between two fixed
frequencies depends on the value of 
$\nu_b$ and also on the shape of the spectrum of the
emitting electrons.
Given the typical shape of the spectrum of
the emitting electrons accelerated during cluster mergers
in our calculations, 
we are able to estimate the minimum typical value of
$\nu_b$ necessary to account for the spectral indices of
the observed radio halos: $\nu_b > 70$ MHz is obtained.
From Eq.~(\ref{rel4}), this limit translates into
a limit on $\chi$ (given in Eq.\ref{Dppt}) :

\begin{equation}
\chi(\eta_t)  \,\, $\gtsim$ \,\, \chi_{min} = 7.4 \times10^{-17}(1+z)^4\bigg(\frac{B_{\mu
G}}{0.5}\bigg)^{-1/2} s^{-1}
\label{chimin}
\end{equation}

\noindent
Radio halos have a typical size $R_H\sim$500 Kpc and they
are found in massive galaxy clusters (M\gtsim$ 10^{15} M_{\odot}$).
Thus we derive the value of $\chi$ for these typical 
clusters in our synthetic population
and find that $\eta_t =0.2-0.3$ is required
to satisfy the condition of Eq.~(\ref{chimin}) during 
major mergers (at $z <0.2$; $B \sim 0.5 \mu$G is adopted).
This is the first important result of our modelling since it 
basically proves that 
if a fraction of the kinetic energy of cluster mergers is
channelled into MS waves  
then this is sufficient to power particle
acceleration in the ICM with the efficiency requested in the
case of radio halos.
Although there are no numerical studies which are 
aimed at a detailed investigation of the cluster turbulence 
injected during merging processes, a general finding of 
high resolution numerical simulations is that a relevant fraction
(10-30 \%) of the thermal energy in galaxy clusters 
is in the form of compressible plasma 
turbulence (e.g., Sunyaev et al. 2003 and ref. therein).
This is in line with 
the requirements of our model.

The second free parameter in our model is $\eta_e$ which gives 
the ratio between the energy injected in relativistic 
electrons during the cluster life and the 
present day thermal energy of the ICM.
The values of $\eta_e$ can be constrained by requiring that 
the model reproduces the typical radio ($L_R$) and hard-X ray 
($L_{HX}$) luminosities observed in galaxy clusters:
$L_R= 10^{40}-10^{41}$erg$s^{-1}$ (Feretti 2003) and
$L_{HX}= 10^{43}-10^{44}$erg$s^{-1}$ (Fusco-Femiano et al.2003).
We derive the requested values for typical massive galaxy
clusters in our synthetic population during the time intervals in which the 
condition of Eq.~(\ref{chimin}) is satisfied.
Making use of Eqs.~(\ref{ke}),~(\ref{syn}), and ~(\ref{IC}) we 
find that $\eta_e= 10^{-4}-10^{-3}$ is sufficient
to match the observed luminosities at $z<0.2$ 
($B \sim 0.5 \mu$G is assumed).
The above $\eta_e$--values are very reasonable
for massive clusters (e.g., Sect.5) and they are
also much smaller than 
those assumed in other modellings of non--thermal emission from
galaxy clusters (e.g., $\eta_e\simeq 0.1$, Sarazin 1999).
This is mainly because in our model the resulting spectrum of the emitting
electrons during an efficient acceleration period is not a simple
power law, but it is peaked at the energies required to emit 
the synchrotron and IC radiation (e.g., Fig.~6) and this
strongly increases the emitting efficiency (see also Sect.~8).

\section{Statistics and Comparison with Observations}

In Sect.~6.2 we have derived a criterion for radio halo
formation: clusters may have radio halos if $\chi(\eta_t)\geq\chi_{min}$.
By making use of this criterion, the goal of this Section is to
calculate the formation probability of radio halos with cluster mass
and to compare expectations with observational constraints.

In order to have a prompt comparison with observations
we calculate the formation probability in the redshift bin $z$=0--0.2 
for three mass bins of the parent clusters $\Delta M$: 
$< 9 \times 10^{14}M_{\odot}$, $9 \times 10^{14}< M <
1.8\times 10^{15}M_{\odot}$, and 
$1.8 \times 10^{15}< M < 3.6 \times 10^{15}M_{\odot}$,
which are consistent with the luminosity bins
adopted to draw the observed statistics 
(Giovannini et al. 1999; Giovannini \& Feretti 2002).

First we run a large number, ${\cal{N}}$, of trees 
for different cluster masses at $z=0$, ranging
from $\sim 10^{14}M_{\odot}$ to $\sim 10^{16}
M_{\odot}$.
Thus, for each $M$, we estimate the formation 
probability 
of radio halos in the mass bin $\Delta M$ as :

\begin{equation}
f_M^{\Delta
M}=\frac{\sum_{j=1}^{{\cal{N}}}t_u^j}{\sum_{j=1}^{{{\cal{N}}}}
(t_u^j+t_d^j)}
\label{partialrate}
\end{equation}

\noindent where $t_u$ is the time that the cluster spends 
at $z<0.2$ in the mass bin $\Delta M$ with
$\chi \geq\chi_{min}$\footnote{Since clusters in our synthetic 
population never have $\chi >> \chi_{min}$, the condition
$\chi \geq\chi_{min}$ guarantees a synchrotron spectral index  
compatible with that of radio halos.} and $t_d$ 
is the time that the same cluster spends in $\Delta M$ with 
$\chi<\chi_{min}$.

Thus the total probability of halo formation in the mass bin $\Delta M$
is obtained by combining all the contributions (Eq.~\ref{partialrate})
weighted with the present day mass function of clusters.

We consider two possible cluster mass functions: 
the Press \& Schecther mass function (1974, P\&S): 

\begin{eqnarray} \label{eq:PSdensity}
n_{\rm PS} (M,z) \, dM =
\sqrt{ \frac{2}{\pi}} \, \frac{ \overline{\rho}}{M}
\, \frac{\delta_{c}(z)}{\sigma^2 (M) } \,
\left| \frac{d \, \sigma (M) }{d \, M} \right|
\times 
\nonumber\\
 \exp \left[- \frac{\delta_{c}^{2}(z)}{2\sigma^{2} (M) } \right]
\, dM
\, ,
\end{eqnarray}

\noindent were $\sigma^2(M)$ is given by Eq.(\ref{sigmam}) and the
other quantities are given in Sect.~3.1, 
and the Sheth \& Tormen (1999, S\&T) mass function
which is obtained from a fit to numerical simulations and 
which predicts smaller and larger values of the cluster 
number density for small and large masses, respectively. 
We checked that the probability to have a radio halo 
obtained by making use of the P\&S and S\&T mass 
functions are consistent within few percent for the 
considered mass bins.

\begin{figure}
\resizebox{\hsize}{!}{\includegraphics{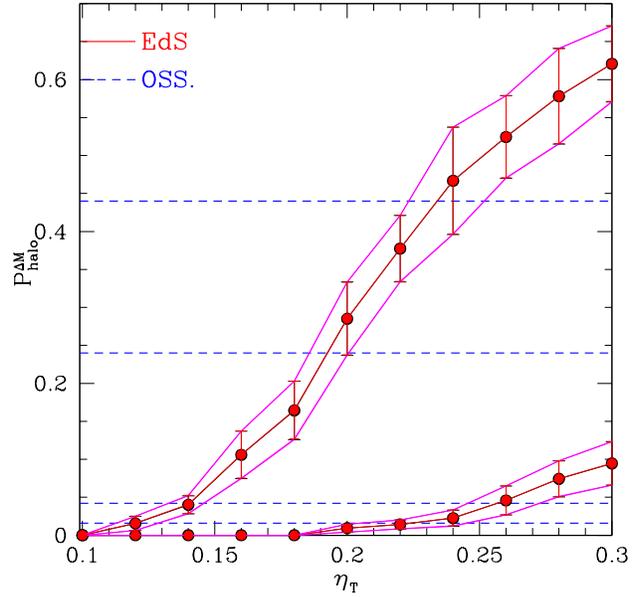}}
\caption[]{Expected formation probability
of radio halos ($R_{H}\simeq 500$ kpc,
$B \sim 0.5 \mu$G) 
in a EdS cosmology as a function of parameter $\eta_t$ in two different 
mass bins (solid lines with error bars): binA$=[1.8-3.6]\,10^{15}M_{\odot}$ 
and binB$=[0.9-1.8]\,10^{15}M_{\odot}$. 
The two bottom dashed lines mark the observed 
probabilities for radio halos in the mass binB while the 
two top dashed lines mark the observed probabilities 
in the mass binA.
The two reported 
observational ranges account for 1$\sigma$ errors.
The theoretical errors are estimated by extracting
sub-samples of galaxy
clusters from the synthetic 
population with a Monte Carlo procedure.}       
\label{prob}
\end{figure}

In Fig.~\ref{prob} we plot the occurrence of
radio halos with a typical radius $R_H \sim 500$ kpc,
as a function of $\eta_t$ compared with
the observated statistics (see caption).
We find that the relatively high occurence of radio halos 
observed in massive clusters can be well reproduced by our modelling
under very reasonable conditions, i.e. that a fraction of 
20-30\% of the energy of the turbulent motions 
(about few percent of the thermal energy) is
in the form of compressible MS waves.
In addition, we find that there is a range of values
of the parameter $\eta_t$
($0.2 \leq \eta_t \leq 0.26$, for $B \sim 0.5 \mu$G) for which 
the theoretical expectations are in agreement with the observed statistics 
in both the considered mass bins: $\sim 30 \%$ and $\sim 4 \%$
in the high and medium mass bins considered, respectively.
Finally, we find that the expected probability 
to form giant radio halos in smaller clusters (not reported in
Fig.~\ref{prob}) is negligible, in agreement with the observations.

\section{Summary and Discussion}

Crucial constraints on the origin of radio halos are
provided by statistical studies which 
show a connection between the formation of these sources 
and cluster mergers, and also find an abrupt increase
of the occurence of radio halos with the mass 
of the parent clusters.

\noindent
{\bf-} The first goal of the present paper is to check if cluster
turbulence generated during mergers may be able to drive efficient 
particle acceleration processes in the ICM.

\noindent
{\bf-} The second goal, in the framework of the turbulent--acceleration 
hypothesis, is to investigate if the hierarchical formation 
process of galaxy clusters can naturally account for the increase 
of the radio halos' occurrence with cluster mass.

To achieve these goals we developed a statistical magneto--turbulent
model which is based on the following steps :

\begin{itemize}
\item[{\it i)}]
Extensive merger trees of galaxy clusters
with different present day masses are obtained.
The trees are calculated making use of a procedure of {\it Binary Merger
Tree Method} which is based on the extended PS formalism (Sect.~3).
The temperature of the ICM is estimated at each redshift
from the observed M--T relationships.

\item[{\it ii)}]
Cluster turbulence is assumed to be injected during cluster
mergers by the crossing of the infalling subclusters into
the larger ones. 
To be conservative, turbulence is assumed to be injected 
in the major subcluster only within the volume sweeped 
by the minor subcluster (Sect.~4).
The injection rate of MS waves is assumed to be 
a fraction, ${\bf\eta_t}$, of the turbulence injection rate. 

The injection spectrum of MS waves is assumed to be 
a simple power law which extends over a broad range of
scales (Sect.~4), or a delta--function from which 
turbulent cascade is originated (Appendix B).
In both cases the maximum/injection scale is fixed
at $L_{inj}\sim 2 r_s$.
The resulting spectrum of MS waves is then calculated assuming 
stationary conditions within a crossing time for each merger and by
taking into account the relevant damping processes 
(or cascading processes) in the ICM (Sect.~4).
The evolution of the spectrum of MS waves during 
cluster formation is calculated by combining the effect of all 
mergers.

\item[{\it iii)}]
The evolution of relativistic electrons in galaxy clusters
is calculated considering the acceleration by MS waves ({\it ii}) 
and the energy losses.
Relativistic electrons are assumed to be continuously
injected in the ICM by shocks, AGNs and star forming galaxies in the
clusters during their life.
The total energy budget injected in the relativistic electrons 
is assumed to be a fraction, ${\bf\eta_e}$, of the thermal energy of 
the clusters at the present epoch.
We do not follow the evolution of the relativistic hadronic 
component since the most important damping of MS waves is
with thermal electrons (Sect.~4.2) and thus the 
relativistic hadrons do not affect significantly the 
electron--acceleration process.
\end{itemize}

To match the redshift range spanned by 
observational studies we calculate the model
expectations for $z < 0.2$.
The comparison between model and 
observations is performed in two main steps :

\begin{itemize}
\item[{\it i)}]
First we consider the case of a typical
massive cluster of our synthetic population
and calculate the expected synchrotron
and inverse Compton emission as a function of
$\eta_t$ and $\eta_e$.
We find that the typical radio luminosity of radio halos
and the HXR luminosities can be obtained by our model 
provided that a fraction of the cluster turbulence, 
$\eta_t \sim [0.2-0.3]\,(B_{\mu G}/0.5)^{-1/2}$ 
($B_{\mu G}$ being the volume averaged
field strength within $R_H$ in units of $\mu$G), is channeled into
MS waves during major mergers and that the energy injected
into relativistic electrons is $10^{-3}-10^{-4}$ times 
the present energy of the thermal pool (Sect.~6.2, see also
the discussion below).

\item[{\it ii)}]
Then, we compute the occurence of radio halos with
the mass of the parent clusters.
More specifically, 
we calculate the minimum particle acceleration
coefficient, $\chi_{min}$, 
which is required 
to efficiently boost the accelerated electron population and
produce radio emission with the spectral slope
typical of radio halos.
We thus identify the galaxy clusters containing a radio halo
as those clusters in our synthetic population for which 
$\chi\geq\chi_{min}$ (see Sect.~7 for details).
The radio halos' occurrence is calculated
in three mass bins 
consistent with those adopted in observational studies
($< 9 \cdot 10^{14}$M$_{\odot}$h$_{50}^{-1}$,
$9 \cdot 10^{14} - 1.8 \cdot 10^{15}$M$_{\odot}$h$_{50}^{-1}$,   
and $1.8 \cdot 10^{15} - 3.6 \cdot 10^{15}$M$_{\odot}$h$_{50}^{-1}$).
We find that for a single range of values
of $\eta_t$ it is possible to account for the 
observed probabilities in all the three mass bins:
about $\sim 30 \%$ and $\sim 4 \%$ in the larger and medium
mass bins, respectively, while the probability to find
a radio halo in a cluster with mass 
$< 9 \cdot 10^{14}$M$_{\odot}$h$_{50}^{-1}$ 
is found to be negligible.
\end{itemize}

As a general conclusion we find that the model expectations are 
in good agreement with the observational
constraints for reliable values of the two free 
model parameters: $\eta_t$, $\eta_e$.

We also find that given these parameters and the physical
conditions in the ICM, the cascade time
of the largest eddies of the MHD turbulence is of the 
order of $\sim$1 Gyr.
Consequently the diffusion and transport of these large
scale eddies and waves may give a fairly uniform turbulent intensity 
within a relatively large volume ($\geq R_H$).
Finally, we find that the two extreme scenarios considered in
our model, i.e. an injection of the MS waves with
a power law spectrum, or with a single scale delta--function, 
provides very similar 
results since the process of particle acceleration
basically depends on the energy flux injected into MS waves
(which is dissipated at collisionless scales) 
and on the physical conditions in the ICM (Appendix B).

Thus, although the necessary approximations adopted in our
formalism, we have shown that particle acceleration processes,  
which are invoked to explain the morphological and
spectral properties of radio halos, can also account for the
statistical properties of this class of objects.

\noindent
The following items need some further discussion:

$\bullet$
An important finding of our calculations is that only massive
clusters can host giant radio halos ($R_H \geq 500$ kpc) and
that the probabilty to form these diffuse radio sources presents 
an abrupt increase for clusters with about 
$M \geq 2 \times 10^{15} M_{\odot}$.

Fig.~4 shows that the energy of the turbulence injected
in galaxy clusters is expected to roughly scale with the
thermal energy of the clusters.
This seems a reasonable finding which immediately
implies that the energy density of the turbulence is an
increasing function of the mass of the clusters, 
${\cal E}_t \propto T \propto M^{a}$.
In addition, in the case of clusters with mass
$M < 10^{15} M_{\odot}$ the infall of subclusters through 
the main one injects turbulence in a volume $V_t$ smaller than
that of giant radio halos, $V_H$, and thus the efficiency of the
mechanism is reduced by about a factor of $V_t/V_H$ (Sect.~4.1).
On the other hand, major mergers between massive subclusters
are expected to inject turbulence on larger volumes,
of the order of $V_H$ (or larger, e.g. Fig.~3b), and thus
the efficiency of the generation of radio halos is not reduced.

More quantitatively, focussing for simplicity
on what happens during a single merger event, 
the efficiency of the particle acceleration in 
the fixed volume $V_H = 4 \pi R_H^3/3$ can be
derived from Eq.(\ref{Dppt}) :
$\chi \propto g(r_s,R_H) (M/R)^{3/2}/\sqrt{T}$,
where the term $T^{-1/2}$ is due to the stronger
damping of MS waves on thermal electrons
with increasing the temperature of the ICM
(Eq.~\ref{dampthms}).
Thus the acceleration efficiency within
$V_H$ is found to scale about with 
$\chi \propto M^{1-a/2} g(r_s,R_H) \propto 
M^{0.75-1.25}$ (0.75 for $M \geq 3 \cdot 
10^{15}M_{\odot}$, 1.25 for $M < 10^{15}M_{\odot}$).

$\bullet$
Several mechanisms can provide injection of turbulence
in the ICM during cluster mergers. 
We have just followed a simple approach which allows us to
estimate the injection of turbulence during the crossing 
of smaller clusters through the more massive ones.
It should be reminded that in the calculations we have adopted 
a typical radius of a radio halo, $R_H \sim 500$ kpc, and assumed that 
turbulence injected in a smaller volume is diffused on the scales of the
radio halo, while the effect of the turbulence injected outside $R_H$ 
is not considered.
However, the stripping radius, in the case of major mergers between very
massive subclusters, can be larger than $R_H \sim 500$ kpc and thus the turbulence
injected by these massive mergers can power particle acceleration
also on larger scales. 
The relativistic electrons accelerated at these scales  
can significantly contribute to the IC spectrum and 
thus the IC luminosities given in this paper (e.g., Fig.~6)
may be underestimated.
On the other hand, the volume integrated synchrotron spectra
should be mainly contributed by the emission produced within $R_H$
due to the expected decrease of the magnetic field strength 
with radius. 
Thus our results, which are essentially
based on the synchrotron properties of radio halos, 
should not be affected by the presence of non--thermal
emission from very large scales.

$\bullet$
An important result of this work is that the energy which is
required to be injected in relativistic electrons 
in the volume of galaxy clusters is of the order of
a few $10^{-4} \times (B_{\mu G}/0.5)^{-2}$ 
of the present day thermal energy of the ICM.
This value basically depends on the 
balance between the electrons' energy
losses and the turbulent--acceleration efficiency 
which is experienced by the relativistic electrons 
injected in the ICM during the last few Gyrs.
Since our calculations are performed 
by assuming the physical conditions of the ICM 
as averaged over the cluster volume, the required 
injected energy in relativistic electrons
may be substantially higher in the central regions 
of the clusters where 
the high density of the thermal plasma 
causes stronger Coulomb losses.
We notice that the required values of $\eta_e$
can be easily provided
by considering the injection of relativistic electrons in the
ICM from AGNs, galactic winds, and large scale shocks
(e.g., Biermann et al., 2003 for a review).

The requested values of the energy injected in 
relativistic electrons in the ICM are calculated
thorugh the paper by assuming 
$s=2.5$ and $p_{\rm min}/mc=60$ (Sect.~5.1).
The results however should not be very sensitive to these
assumptions, and they would be only sensitive to the
total number of relativistic electrons injected in the
ICM during the cluster life.
Indeed, the turbulence experienced in the ICM basically
increases the cooling time of the injected electrons which are then
mantained at the peak of their {\it cooling--time curve} (i.e., at
$\gamma \sim 100-200$, e.g. Sarazin 1999) and thus boosted at
higher energies during an efficient re--acceleration period.
In order to test the poor dependence of our results on
the assumptions on $s$ and $p_{\rm min}$, 
we re--calculate the value of $\eta_e$ by assuming 
$s=2.2-3.0$ and $p_{\rm min}/mc=20-100$.
We find that different assumptions requires values
of $\eta_e$ within a factor of $\sim 3$ to 
reproduce a given synchrotron power.
In particular we find that $\eta_e$ decreases with increasing $s$ 
(or with decreasing $p_{\rm min}$).

It should be stressed that the amount of
injection of relativistic electrons required by our model is 
orders of magnitude smaller than that needed by models which 
assume a simple continuous injection of a 
power law energy distribution of the emitting
electrons in the ICM (e.g., Sarazin 1999).
This is mainly because during an 
efficient acceleration period the spectrum of the 
relativistic electrons is not a steep power law in which
the bulk of the electrons is at low energies.
During this period the bulk of the electrons, accumulated
at $\gamma \sim 100-300$, is boosted 
at higher energies and essentially piled up in the energy 
range responsible for the synchrotron emission in the radio band.

$\bullet$
Since the present work is not aimed at reproducing in detail the 
properties of radio halos, in our calculations we assume 
that the magnetic field strength (within $R_H$) is roughly 
constant ($B \sim 0.5 \mu$G is assumed to constrain $\eta_t$
and $\eta_e$).
Larger values of $B$ (but still under the conditions in which 
the radiative cooling of electrons is dominated by IC emission,
i.e. $B < 3 \mu$G) would allow to radiate the synchrotron photons 
at higher frequencies (Sect.~6.3, Eq.~\ref{chimin}) and this would imply 
that lower values of $\eta_t$ ($\eta_t \propto B^{-1/2}$, 
Sect.~6.2) are required to form radio halos.
On the other hand, the discovery of HXR tails in 
galaxy clusters has revealed that the non--thermal spectra
of these objects are dominated by the IC component which
has a luminosity $\sim 10^3$ times larger than the synchrotron
component. If confirmed, these observations indicate that the 
volume--averaged magnetic field strength should be $< 0.5 \mu$G 
(e.g., Fusco--Femiano et al., 2003).
However, as discussed above, a relevant contribution to the 
IC spectrum of galaxy clusters can be provided by electrons accelerated 
by turbulence injected in the outer regions ($\geq R_H$) and thus values 
$B$\gtsim$1 \mu$G in the synchrotron emitting volume may be still 
compatible with the observed IC components.

$\bullet$
As already stated in the model calculations we have assumed 
a typical mean radius of the radio halos and a value of
the magnetic field strength $B$ which are independent 
from the mass of the parent clusters.
If radio halos in more massive clusters are larger than those 
in smaller ones, then this approach should 
underproduce the expected probability to find radio halos in 
the smaller clusters with respect to the larger ones.
The fact that the values of $\eta_t$ required to match observations
in the intermediate mass bin are found to be
slightly larger than those in the more massive bin (Figs.7 and A1)
may reflect this effect.
On the other hand, if $B$ increases with the mass of the parent clusters
(with $B \leq 3 \mu$G) then the synchrotron
emission would be boosted at higher frequencies and the
expected probability to find radio halos in the case of
larger clusters would be slightly
increased with respect to our
present expectations.

$\bullet$
Finally, the model results obtained with a EdS cosmology have been 
compared with those obtained assuming a 
$\Lambda$CDM cosmology (Appendix A).
We find that the possibility to explain the observations
in the redshift bin $0-0.2$ 
does not depend critically on the adopted cosmology.
In particular, although the model is found to be slightly 
less efficient in a $\Lambda$CDM cosmology,
also in this case the occurence of radio halos can be matched for
viable values of the parameters, and a single range of $\eta_t$
is found to be able to explain observations in all
the mass bins; this would strengthen our conclusions.

Future studies with radio (LOFAR, SKA) and hard X--ray (ASTRO-E2, NEXT) 
observatories will be crucial to constrain the occurrence and evolution 
of the observed non--thermal diffuse emission in galaxy clusters and thus
to perform a detailed comparison between observations and model
expectations.

\section{Acknowledgements}
We are indebted to G.Setti for a careful reading of the manuscript 
and for useful comments. 
We thank P.Blasi, K.Dolag, L.Moscardini and G.Tormen 
for useful discussions.
We thank the referee for useful comments and suggestions which have
improved the presentation of this paper.
G.B. and R.C. acknowledge partial support from CNR grant
CNRG00CF0A, G.B. acknowledge partial support from INAF through
grant D4/03/15.

\appendix 
\section{The case of $\Lambda$CDM model}

\noindent
In order to have a prompt comparison with observational
studies, all the results given in the paper are obtained in a
Einstein-De Sitter (EdS) cosmology.
In this Appendix we adopt a $\Lambda$CDM 
cosmology 
(we use $\Omega_m(0)=0.3$, $\Omega_{\Lambda}(0)=0.7$, 
$\sigma_{8}=0.9$ and $h_{0}=0.7$)
and discuss the differences with the EdS case.

In the $\Lambda$CDM cosmology the 
critical overdensity (in Eq.\ref{mergermerger})
as a function of 
the cosmic time is given by (Kitayama \& Suto 1996):

\begin{equation}
\delta_{c}(t)
=\frac{D(t_{0})}{D(t)}\bigg(1+0.0123log\Omega_{m}(z)\bigg)
\label{delc}
\end{equation}

\noindent 
where $\Omega_{m}(z)$ is the mass density ratio at the redshift z,

\begin{equation}
\Omega_{m}(z)
=\frac{\Omega_m(0)(1+z)^3}{\Omega_m(0)(1+z)^3+\Omega_{\Lambda}(0)}
\label{Omz}
\end{equation}
The growth factor in Eq.(\ref{delc}) 
is (Peebles 1980, Eq.13.6):

\begin{equation}
D(x)
=\frac{(x^{3}+2)^{1/2}}{x^{3/2}}\int_{0}^{x}y^{3/2}(y^3+2)^{-3/2}dy
\label{Dx}
\end{equation}

\noindent where $x_{0}\equiv(\frac{2\Omega_{\Lambda}}{\Omega_{0}})^{1/3}$ 
and $x=x_{0}/(1+z)$.

\noindent 
The ratio of the average density of the cluster to the mean density 
of the universe at a given z, $\Delta_{c}(z)$ (in Eq.(\ref{Rv})), 
in the $\Lambda$CDM model
is given by (Kitayama \& Suto 1996):

\begin{equation}
\Delta_{c}(z)=
18\pi^2(1+0.4093\omega(z)^{0.9052})
\label{Dc}
\end{equation}

\noindent 
where $\omega(z)\equiv \Omega_{m}(z)^{-1}-1$.

\begin{figure}
\resizebox{\hsize}{!}{\includegraphics{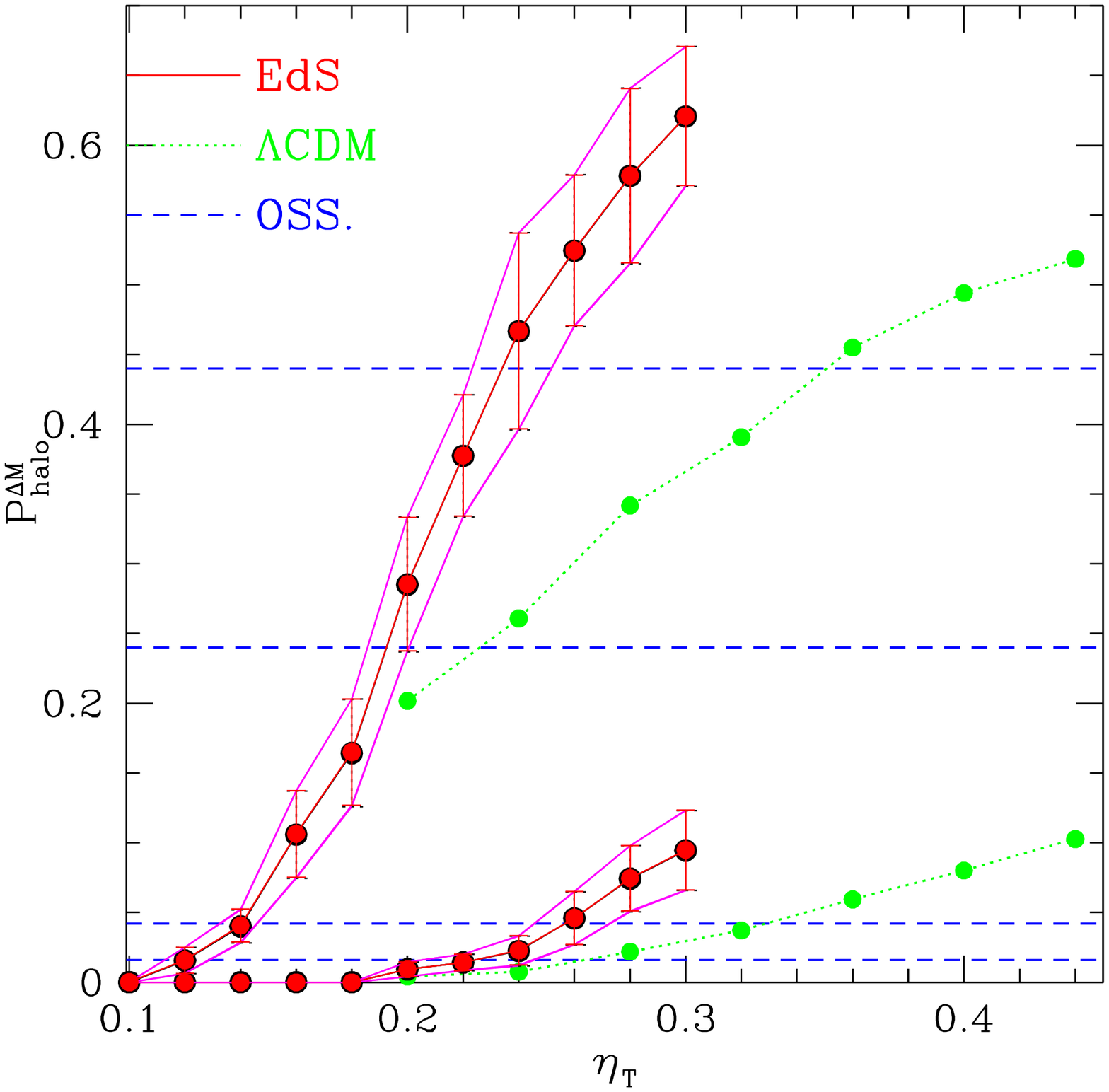}}
\caption[]{Expected formation probability of radio halos 
($R_{H}\simeq 500 h_{50}^{-1}$kpc, $B \sim 0.5 \mu$G) 
as a function of parameter $\eta_t$ in a EdS cosmology 
(solid lines with error bars) and in a $\Lambda$CDM cosmology 
(dotted lines) in the mass bins:
binA=$[1.8-3.6]\,10^{15}M_{\odot}\,h_{50}^{-1}$ and 
binB=$[0.9-1.8]\,10^{15}M_{\odot}\,h_{50}^{-1}$ for EdS case
and  binA=$[1.9 - 3.8] \cdot 10^{15}$ M$_{\odot}$h$_{70}^{-1}$ and 
binB=$[0.945 - 1.9] \cdot 10^{15}$ M$_{\odot}$h$_{70}^{-1}$ 
for the $\Lambda$CDM model.
The two bottom dashed lines mark the observed 
probabilities for radio halos in the mass binB while the 
two top dashed lines mark the observed probabilities 
in the mass binA; observational regions account for 1$\sigma$ errors.}
\label{prob_lambda}
\end{figure}

Following the procedures adopted in the case of the 
EdS cosmology, we compute merger trees
(Sect.~3), turbulence injection rate and spectrum of the MS waves
(Sect.~4), particle evolution (Sect.~5) and non-thermal emission
(Sect.~6) from galaxy clusters and thus the expected formation 
probability of radio halos for $z<0.2$ (Sect.~7).
In Fig.\ref{prob_lambda}, we report the comparison between the
probability to form radio halos obtained in the two cosmologies.
The comparison is derived by converting the virial mass of the
clusters from a EdS into a $\Lambda$CDM model:

\begin{equation}
M_v^{\Lambda}=M_v^{EdS}\times\bigg({{[\Delta_c(t)\rho_m(t)]_{EdS}}\over
{[\Delta_c(t)\rho_m(t)]_\Lambda}}\bigg)^{1/2}
\label{M_lEdS}
\end{equation}

\noindent
where $\rho_m$ is the mean mass density (dark and barionic)
of the Universe.
Thus the calculations with a $\Lambda$CDM model are performed for the 
mass bins $[0.945 - 1.9] \cdot 10^{15}$ M$_{\odot}$h$_{70}^{-1}$ and 
$[1.9 - 3.8] \cdot 10^{15}$ M$_{\odot}$h$_{70}^{-1}$.

As expected, we find that at $z<0.2$ the results are 
relatively independent from the considered cosmology,
with the $\Lambda$CDM model being only slightly less efficient. 
In particular, as in the EdS case we note that it is possible
to find a unique interval in $\eta_t$ in which the model reproduces
the observed halo formation probability for both the cluster-mass bins.

In the $\Lambda$CDM Universe the structures 
start to grow at early time with respect to the EdS case,
the merging rate at $z<0.2$ is consequently reduced,
and thus particle acceleration is less efficient.
However, this is roughly compensated by the fact that in a $\Lambda$CDM
Universe the observed radio halos are smaller and less luminous than
in the EdS case.

\section{Turbulence injection at a single scale}

In this Appendix we adopt the scenario in which MHD turbulence
is injected in the ICM at a large single--scale, 
$k_{min} \sim \pi/r_s$, 
from which the MHD turbulence cascade is originated.

The mean free path, $L_{mfp}$, in the ICM marks the 
boundary between the collisionless regime ($k > 2 \pi/L_{mfp}$)
and the collisional regime ($k < 2 \pi/L_{mfp}$).
It is given by (e.g., Braginskii 1965):

\begin{equation}
L_{mfp}({\rm kpc}) \sim 300
( {{ T }\over{ 10^8 }} )^2
( {{ n_{th} }\over{10^{-4}}} )^{-1}
\label{mfp}
\end{equation}

\noindent
which, for the typical values of the cluster temperatures and 
of the mean thermal density within $V_H$, is of the
order of 100 kpc.

Once MS waves are injected at $k_{min}$, the process of wave--wave 
coupling generates a turbulence cascade.
The cascade time of fast MS waves at the wavenumber $k$
is given by (e.g., Yan \& Lazarian 2004):

\begin{equation}
\tau_{kk}(k) \sim {{ v_{M} }\over{k v_k^2}} \sim 
{{ v_{M} \rho }\over{k^2 W_k}}
\label{tkk}
\end{equation}

\noindent
so that the diffusion coefficient in Eq.(\ref{turbulencems}) 
is given by :

\begin{equation}
D_{kk} \sim {{k^2}\over{\tau_{kk}}}
\sim {{ v_A^2 }\over{v_{M}}}
{{ k^4 W_k}\over{2 W_B}}
\label{dkk}
\end{equation}

\noindent
In the quasi linear regime, the spectrum of the waves due to 
the cascading process can be calculated solving Eq.(\ref{turbulencems}) 
and neglecting the contribution due to the damping terms :

\begin{equation}
{{\partial W_{\rm k}(t)}\over
{\partial t}} =
{{\partial}\over{\partial k}}
\left( D_{\rm kk} {{\partial W_{\rm k}(t)}\over{\partial k}}
\right) +
I_{\rm k}
\label{turbulencems_app}
\end{equation}

\noindent
with $I_{\rm k} = I_o \delta(k-k_{min})$ and
$I_o \simeq \eta_t v_i^3 \rho (\pi r_s^2/V_H)$ (Sect.~4.1).
The stady state solution of Eq.(\ref{turbulencems_app}) 
is a Kraichnan--like spectrum :

\begin{equation}
W_{\rm k} \simeq \left( {{2 \rho I_o v_{M}}\over{3}}
\right)^{1/2} k^{-3/2}
\label{sstate}
\end{equation}

\noindent
This spectrum extends down to a truncation scale at which the
cascading time, $\tau_{kk}$, becomes substantially
larger (i.e., $\xi$ times, $\xi \sim 1-3$) than the damping time
scale, $\tau_d \sim \Gamma_{th,e}^{-1}$ (Eq.~\ref{dampthms}).
In the collisionless regime, this truncation scale, 
$L_{tr}\sim 2\pi/k_{tr}$, is obtained from Eqs.(\ref{dampthms}), 
(\ref{dkk}), and (\ref{sstate}), one has :

\begin{equation}
L_{tr} \simeq {{ 0.23 }\over{ \xi^2 \eta_t}} 
( {{T}\over{10^8}} )^{3/2}
( {{ v_{i} }\over{ 10^3 {\rm km/s} }} )^{-3}
\left(
{4\over 3} {{ R_H^3}\over{r_s^2}} \right)
\label{ltr}
\end{equation}

\noindent
which typically falls in the range 10--30 kpc for our 
synthetic clusters (note that such scale is
smaller than $L_{mfp}$ and thus 
the estimate can be done under 
the assumption of a collisionless
regime), i.e. a factor of 30--100 smaller 
than the value of the typical turbulence
injection scale (Sect.~4.1-4.2, Fig.~3b).

The picture of the model in this Appendix 
is thus that the injection of MHD turbulence 
occurs at a maximum scale of the order of 1 Mpc which is larger
but relatively close to the scales typical of the collisionless regime.
The wave-wave coupling then leads to a power-law inertial
range with a Kraichnan spectrum which is approximatively mantained 
down to $\sim 10-30$ kpc.
At these scales the damping time with the thermal electrons
becomes considerably shorter than the cascading time--scale
and the turbulence cascade is broken.

Under these conditions the acceleration time of 
relativistic electrons, $\tau_{\rm acc}$,
is dominated by the contribution from the spectrum of the 
waves at the truncation scale and it 
can be obtained from Eqs.(\ref{dppms})
and (\ref{chi}) :

\begin{equation}
\tau_{\rm acc}^{-1} = \chi 
\propto 
v_M^2 W_k(k=k_{tr}) k_{tr}^2
\label{accB}
\end{equation}

An important point is to check if the scenario
adopted in Sect.~4 and that adopted in this Appendix 
give consistent results. In the scenario adopted in Sect.~4.2,
the spectrum of the MS waves is approximately
given by :

\begin{equation}
W_k \sim I_k \tau_d(k)
\label{app1}
\end{equation}

\noindent
and thus, since $I_o = \int I_k dk$ and the damping time scale
is $\tau_d(k) \sim \Gamma_{e,th}^{-1} \propto k^{-1}$, one has :

\begin{equation}
D_{pp} \sim c_{pp}
\int W_k k dk
\sim I_o \tau_d(k_{tr}) k_{tr}
\label{app2}
\end{equation}

\noindent
where $c_{pp}$ does not depend on the turbulence spectrum
and energy (Eq.~\ref{dppms}).
On the other hand, in the scenario adopted in this
Appendix the spectrum of the MS waves is approximately
given by :

\begin{equation}
W_k \sim {{I_o}\over{k}} \tau_{kk}
\label{app3}
\end{equation}

\noindent
and thus, since the cascading time scale
is $\tau_{kk} \propto k^{-1/2}$, one has :

\begin{equation}
D_{pp} \sim c_{pp}
\int W_k k dk
\sim 2 \, I_o \tau_{kk}(k=k_{tr}) k_{tr}
\label{app4}
\end{equation}

\noindent
as a consequence, since $k_{tr}$ is the scale at which the damping
time scale and the cascading time scale are comparable, 
we expect that the two scenarios would 
provide a similar acceleration efficiency.

\noindent
More specifically, we can calculate the electron acceleration
coefficient due to a single merger event (with $r_s \geq R_H$)
in the framework of the scenario adopted in this Appendix.
From Eqs.(\ref{sstate}), (\ref{ltr}), (\ref{accB}), (\ref{vi}),
and the expression for $I_o$ given in this Appendix, one finds :

\begin{equation}
\chi \sim 2 \times 10^{-16} \xi \eta_t 
\left( {{ M}\over{2 \times 10^{15} M_{\odot} }} \right)^{3/2}
\left( {{kT}\over{7\,keV}} \right)^{-1/2}
\left[ {{(r_s/500 kpc)^2}\over{(R_H^3/500 kpc)}} \right]
\label{chiApp}
\end{equation}

\noindent
which is close to the value given in Eq.(\ref{Dppt}),
and thus proves that the main results of our model 
do not crucially depend on the assumptions
on the specific injection process (and spectrum) 
of the MS waves.

\section{Synchrotron and IC Radiation}

In this Appendix we briefly describe the formalism adopted to
calculate the synchrotron and IC emission from 
galaxy clusters.

The synchrotron emissivity is given by:

\begin{equation}
J_{syn}(\nu,t)=\frac{\sqrt{3} e^3 B}{mc^2}\int_p\int_0^{\pi/2} dp d\theta
sin^2 \theta N(p,t) F(\frac{\nu}{\nu_c})
\label{syn}
\end{equation}

\noindent where $F(\frac{\nu}{\nu_c})$ is the synchrotron Kernel given
by (e.g., Rybicki \& Lightman, 1979):

\begin{equation}
F(\frac{\nu}{\nu_c})=\frac{\nu}{\nu_c}\int_{\nu/\nu_c}^{\infty}
K_{\frac{5}{3}}(y)dy
\label{kernel}
\end{equation}

\noindent with $\nu_c=(3/4\pi) p^2 e B sin\theta/(mc)^3$.\\

The IC emissivity due to the scattering off the CMB photons by 
relativistic electrons is given by (e.g., Blumenthal \& Gould,1970):

\begin{eqnarray}
\lefteqn{J_{IC}(\nu_1,t)=(2\pi r_om^2c)^2h\nu_1^2\int_{\nu}\int_p
d\nu dp\frac{N(p,t)p^{-4}}{exp(\frac{h\nu}{k_BT_z})-1}\times {} }
\nonumber\\
& & {}(1+2\ln(\frac{\nu_1m^2c^2}{4p^2\nu})+\frac{4p^2\nu}{m^2c^2\nu_1}
-\frac{m^2c^2\nu_1}{2p^2\nu})
\label{IC}
\end{eqnarray}

\noindent 
where $T_z=2.73(1+z)$ is the temperature of the CMB photons.

\end{document}